\documentclass[12pt]{iopart}
\usepackage{iopams}
\usepackage{epsfig}
\usepackage{graphicx}

\newtheorem{theorem}{Theorem}[section]

\newtheorem{definition}[theorem]{Definition}
\newtheorem{lemma}[theorem]{Lemma}

\newtheorem{remark}[theorem]{Remark}
\newtheorem{reminder}[theorem]{Reminder}

\begin{document}
\title[Interacting particle systems, KPZ universality, and random matrices]{A pedestrian's view on interacting particle systems, KPZ universality, and random matrices}
\author{Thomas Kriecherbauer}
\address{Fakult\"at f\"ur Mathematik, Ruhr-Universit\"at Bochum}
\author{Joachim Krug}
\address{Institut f\"ur Theoretische Physik, Universit\"at zu K\"oln}
\eads{\mailto{thomas.kriecherbauer@ruhr-uni-bochum.de} and \mailto{krug@thp.uni-koeln.de}}
\date{today}

\begin{abstract}
These notes are based on lectures delivered by the authors at a Langeoog seminar
of SFB/TR12 \textit{Symmetries and universality in mesoscopic systems}  to a mixed audience of mathematicians and theoretical physicists. After a brief outline of
the basic physical concepts of equilibrium and nonequilibrium states, the one-dimensional
simple exclusion process is introduced as a paradigmatic nonequilibrium interacting particle system.
The stationary measure on the ring is derived and the idea of the
hydrodynamic limit is sketched. We then introduce the phenomenological
Kardar-Parisi-Zhang (KPZ) equation and explain the associated universality
conjecture for surface fluctuations in growth models. This is followed
by a detailed exposition of a seminal paper of Johansson \cite{Jo1}
that relates the current 
fluctuations of the totally asymmetric simple exclusion process (TASEP) to the
Tracy-Widom distribution of random matrix theory. The implications of
this result are discussed within the framework of the KPZ conjecture.
\end{abstract}

% \maketitle
%related exact results and discuss them in the context of KPZ theory.
\section{Introduction and outline}

In statistical mechanics the study of systems that are far from
equilibrium continues to attract considerable attention both in the
physics and in the mathematics literature. As it turns out,
exclusion processes first introduced by Spitzer
\cite{Spitzer1970} and their generalisations provide an excellent
set of models that display rich and interesting nonequilibrium phenomena. Moreover,
these processes are intimately related to a number of different
models from statistical mechanics, combinatorics, probability and
random matrix theory and a fruitful interplay between these fields,
triggered by the seminal work of Baik, Deift and Johansson
\cite{Baik99}, continues to produce spectacular results that yield
very precise information on the behavior of such systems.

It is beyond the scope of this article to explain all or even a fair amount
of these results in detail. Instead, we remain faithful to the original goal
of the set of lectures on which this manuscript is based, namely to explain
the fascinating developments in this field to an audience of scientists working
in many different areas of mathematics and theoretical physics. We do not assume
any significant acquaintance with the concepts of statistical mechanics or
probability theory. In order to keep our presentation as elementary as possible
we shall always focus on the simplest cases. Readers who wish to obtain more
information on a specific topic will be referred to the literature.
Here we will make use of the numerous reviews that have appeared recently in this area.

In the remaining part of the introduction we provide a short outline
of the topics that will be treated in this paper.

In Sect.~\ref{Equilibrium} we begin our discussion by explaining in
general terms what physicists mean by the distinction between
equilibrium and nonequilibrium systems, and by describing different
types of nonequilibrium behavior. We then introduce in
Sect.~\ref{Exclusion} a class of stochastic models, known as
\textit{simple exclusion processes}. They describe the stochastic motion of
interacting particles on a lattice where the interaction is given by
the exclusion property, i.e. two particles may not occupy the same
site simultaneously. Interacting particle systems provide useful
models for various nonequilibrium phenomena. Technically, they are
(discrete or continuous time) Markov chains, and we will argue below
in Sect.~\ref{Markov} that a simple yet precise criterion for the
equilibrium vs. nonequilibrium character of a given system can be
formulated within the general theory of Markov chains
\cite{Zia2007}.

We then proceed to explain in more detail how one dimensional simple
exclusion processes can be analyzed, specializing to the case
where particles move only to
neighboring lattice sites. If the probabilities to
move right or left differ from each other one obtains the
\textit{asymmetric simple exclusion process} (ASEP). Mostly this paper
will be concerned with ASEP and its subcase TASEP (\textit{totally
asymmetric simple exclusion process}) where all motion is
unidirectional. The criterion of Sect.~\ref{Markov} identifies ASEP (and
consequently TASEP) as nonequilibrium systems. In fact, ASEP has
become a paradigmatic model for driven transport of a single
conserved quantity and most of our discussion is focused on this
class of models. For readers who are not familiar with ASEP and TASEP
it might be useful at this point to have a look at the precise
definition in Sect.~\ref{jsec3.1}.

We begin our exposition of the analysis of ASEP with a discussion of
stationary measures in Sect.~\ref{Measures} where it is explained
why the uniform distribution always provides a stationary measure in
the simplest case of periodic boundary conditions.

For a macroscopic description of the dynamics the notion of the
hydrodynamic limit is introduced. The time evolution of the
macroscopic particle density is then described by a hyperbolic PDE
that can be solved by the method of characteristics
(Sect.~\ref{Hydro}). After this has been established it is most
natural to ask how much the process fluctuates around this
macroscopic description. At this point it is useful to realize that
ASEP is equivalent to a specific random model for surface growth
which is known as 'corner growth' or 'single step' model
(Sect.~\ref{Growthmodel}). Kardar,
Parisi and Zhang (KPZ) conjectured in their seminal paper
\cite{Kardar1986} that the fluctuation properties of a large class
of (growth) models are universal. Assuming that KPZ universality can
be applied to this particular growth model, one may obtain
predictions for the scaling exponents of the fluctuations of the
height of the surface and of the scaling exponent of the (spatial)
correlation length. The KPZ conjecture and its implications for ASEP
will be explained in Sect.~\ref{KPZ_conjecture}.

The KPZ conjecture is based on the KPZ equation which essentially
adds stochastic driving to the hyperbolic PDE that describes the
hydrodynamic limit. Unfortunately, the KPZ equation is difficult to
analyze and it was only very recently that fluctuation results for
solutions of the KPZ equation became available (cf.
Sect.~\ref{T8.4}). Before that an alternative approach to studying fluctuation properties was taken that turned out to be very fruitful. Rather than dealing with the KPZ equation itself one analyzes various specific models that are believed to belong to the
KPZ universality class. A major breakthrough was achieved through a
spectacular discovery on Ulam's problem for random permutations.

In the early 60's Ulam raised the question to determine the asymptotics of the length of the longest increasing subsequence of a random permutation of the numbers $1, \ldots,
N$, where it is assumed that all $N!$ permutations are equally
likely. It took about 15 years to prove that the expected value of the
length of the longest increasing subsequence behaves like
$2\sqrt{N}$ as $N$ becomes large and there was strong numerical
evidence that the fluctuations around the mean are of order
$N^{1/6}$. In a remarkable paper Baik, Deift, and
Johansson \cite{Baik99} did not only prove that $N^{1/6}$ was indeed the correct
scaling, but they also identified the limiting distribution for the
appropriately rescaled fluctuations. It came as a surprise that
the limiting distribution coincides with the Tracy-Widom
distribution that describes the fluctuations of the largest
eigenvalue of matrices from the Gaussian Unitary Ensemble (GUE) as
the matrix size tends to infinity.

The article \cite{Baik99} became the starting signal for an explosion of
research activities that continue until today. It became immediately
clear that there is a variety of related combinatorial models (e.g.
growth models, last passage percolation models, tilings, directed
polymers in a random environment, tandem queues) that can be
analyzed at the same level of detail and where Random Matrix
distributions appear in the asymptotic description. For further
information we refer the reader to the surveys 
\cite{Aldous1999,TTracyWidom02,TJohansson02,TKonig05,Spohn2006,Majumdar2007}
and the monograph \cite{TBDS}. Two results that are of particular relevance for our
discussion of KPZ universality were obtained independently by
Pr\"ahofer and Spohn \cite{Prahofer00b} and Johansson \cite{Jo1}.

Pr\"ahofer and Spohn used the results in \cite{Baik99}, and further
developments in \cite{TBaRa01}, to describe the height fluctuations
of the polynuclear growth model (which is somewhat different from
the corner growth/ single step model mentioned above, cf.
Sect.~\ref{Universality}). They obtained the scaling law predicted
by KPZ universality. Moreover, they were able to identify the
limiting distributions of the fluctuations. Again these are given by
the Tracy-Widom distributions of Random Matrix Theory. A distinction
needs to be made depending on the curvature of the surface. Flat
surfaces lead to statistics from the Gaussian Orthogonal Ensemble
(GOE), whereas the height-fluctuations of curved surfaces are
described by GUE statistics.

On the other hand, the results that Johansson presented in
\cite{Jo1} immediately apply to the totally asymmetric simple
exclusion process with step initial condition. One obtains the
scaling exponent and the limiting distribution for the fluctuations
of the particle flux, that are again given by the GUE Tracy-Widom
distribution. We devote Sects.~\ref{tsec:R} -- \ref{tsec:A} to
explain \cite{Jo1} in great detail. This part of our presentation can be viewed as an expanded and more self-contained version of Sects.~3 and 4 of the review \cite{Sas1} by T. Sasamoto.

According to the philosophy of our paper we explain the results of
Johansson in the simplest case. More precisely, we consider the
particle flux at the origin for a discrete time version of TASEP
(dTASEP, cf. Sect.~\ref{jsec3.1} (iii)) with step initial data. In
Sect.~\ref{tsec:R} we formulate Johansson's result in Theorem
\ref{tsatz:R.1} and discuss its relation to KPZ universality.

The proof of Theorem \ref{tsatz:R.1} naturally falls into two parts.
The first part is of combinatorial nature. Via a
representation by waiting times (Sect.~\ref{tsec:C.1}) the problem
is mapped to finding the longest subsequence in a list of
alphabetically ordered two-letter random words that is weakly
increasing in the second letter (Sect.~\ref{tsec:C.2}). By the
Robinson-Schensted-Knuth algorithm (Sect.~\ref{tsec:C.3}) one may
represent the random words by pairs of Semi Standard Young Tableaux
of the same shape (see Definition \ref{tdef:C.1}). The advantage of
this representation is twofold: On the one hand the length of the
longest weakly increasing subsequence is simply given by the length
of the first row of the corresponding Young Tableau. On the other
hand there exist explicit formulae for counting the number of
Semistandard Young Tableaux of a given shape, that can be derived
using Schur polynomials (Sect.~\ref{tsec:C.4}). The result of all
this reasoning is formula (\ref{teq:C.28}), where $\Delta$ denotes
the Vandermonde determinant (see also (\ref{teq:R.9}) and
Definitions \ref{tdef:R.1} and \ref{tdef:R.2}).

The second part of the proof of Theorem \ref{tsatz:R.1} is the
asymptotic analysis of (\ref{teq:C.28}). The key observation is that
the right hand side of (\ref{teq:C.28}) has exactly the same
structure as the formula for the distribution of the largest
eigenvalue of GUE matrices. In particular the method of orthogonal
polynomials (Sect.~\ref{tsec:A.1}) can be applied to complete the
proof of Theorem \ref{tsatz:R.1} in Sect.~\ref{tsec:A.2}. The
somewhat miraculous appearance of the Tracy-Widom distribution for
the fluctuation of the particle flux of dTASEP is now explained on a
technical level by the fact that Hermite polynomials (used for GUE)
and Meixner polynomials (used for dTASEP) look the same near their
respective largest zeros after appropriate rescaling. The similarity
of Hermite and Meixner polynomials is no coincidence. We briefly
discuss the universal behavior of orthogonal polynomials in
Sect.~\ref{tsec:A.3}.

As it was mentioned above the work of Johansson \cite{Jo1} and of
Pr\"ahofer and Spohn \cite{Prahofer00b} mark the beginning of a
broad stream of research activities that continues to produce new
and exciting results at a rapid pace. In Sect.~\ref{Universality} we
briefly sketch and summarize those directions of recent research
that are closely related to the question of KPZ-universality. The
first generalization beyond \cite{Jo1} that we describe concerns the initial conditions.
The results of Johansson apply for step initial conditions where
every site to the left of the origin is occupied whereas every site
to the right is empty. Based on their work on the polynuclear growth
model Pr\"ahofer and Spohn \cite{Prahofer01} formulated a conjecture
for the fluctuations of the flux for TASEP in the case of a general
initial step profile with arbitrary constant particle densities
$\rho_L$ and $\rho_R$ to the left resp. right of the origin. In
Sect.~\ref{T8.1} we explain this conjecture that has recently been
fully established by Ben Arous and Corwin \cite{TBeCo09}. Most
remarkably, in a series of papers
\cite{TTracyWidom08b}-\cite{TTracyWidom10} C. Tracy and H. Widom
were able to extend some of these results to general ASEP. It should
be pointed out that their proof is based on the Bethe Ansatz and
does not use any of the nice but very special combinatorial identities that were crucial in the argument of Johansson.

Furthermore we provide in  
Sect.~\ref{Universality} pointers to the recent literature regarding spatio-temporal correlations for (T)ASEP (Sect.~\ref{T8.2}), interacting particle systems beyond (T)ASEP
(Sect.~\ref{T8.3}), fluctuation results for the KPZ equation
(Sect.~\ref{T8.4}), and physical experiments where KPZ behavior can
be observed (Sect.~\ref{T8.5}). We conclude the paper with a few
remarks on integrability and universality.

\section{Equilibrium and nonequilibrium states}
\label{Equilibrium}

The most fundamental concept of statistical physics is the distinction between
\textit{microstates} and \textit{macrostates} in the description of systems
with many degrees of freedom. To fix ideas, consider a classical 
$N$-particle system (say, a gas in a box) 
described by a Hamilton function $H(q,p)$ of position
variables $q = (q_1,...,q_{dN})$ and momenta $p = (p_1,...,p_{dN})$. 
Particles move in a region $\Omega \subset \mathbb{R}^d$ of volume $V = \vert \Omega
\vert$. Then a \textit{microstate} is simply a point $(q,p)$ in phase space, whereas
a \textit{macrostate} will be defined for the purposes of these lectures as 
a measure $P_X(q,p) dq \, dp$ parameterized by a set of \textit{macroscopic}
state variables (in short \textit{macrovariables}) $X$. 
Here $P_X(q,p)$ is a function on phase space and $dq \, dp$
denotes the canonical Liouville measure. 

Examples of macrovariables
are energy, density, temperature or pressure. The macrovariables
parametrizing the macrostate $P_X$ could have a dependence on space and time, 
but to be useful they should be chosen such that they are slowly varying. 
This singles out in particular
the conserved quantities of the underlying $N$-particle system as candidates for
macrovariables. The mapping from the microstate $(q,p)$ to the 
macrovariables $X$ is many-to-one, and the measure 
$P_X(q,p) dq \, dp$ gives the probability to find the system in a particular
set of microstates $(q,p)$ under the constraint that the macroscopic state is described
by $X$. In principle, the time dependence (if any) of $P_X(q,p)$ is induced by the classical
Hamiltonian dynamics of the microstate variables $(q,p)$, but in practice
well-chosen macrovariables are often found to satisfy autonomous evolution
laws, such as the equations of hydrodynamics. The derivation of macroscopic evolution 
equations from microscopic Hamiltonian dynamics is the goal of \textit{kinetic theory}.
A (much simplified) version of this problem will be addressed below in 
Sect.~\ref{Hydro}.  
 
In this perspective, \textit{equilibrium states} are a subclass of macrostates
which are attained at long times by a system that is isolated or in contact
with a time-independent, spatially uniform environment. Characteristic properties
of equilibrium states are that
\begin{itemize}

\item the macrovariables $X$ are time-independent and spatially homogeneous, and 

\item there are no macroscopic currents (e.g., of mass or energy). 

\end{itemize}
The two most important examples of equilibrium states are the following:
\begin{itemize}

\item[a.)] In an \textit{isolated} system the energy $E$ is conserved, 
the appropriate macrovariables are $X = (E, V, N)$ and the equilibrium
state is the measure induced by the Liouville measure 
on the energy shell $\{(q,p): H(q,p) = E\}$.
This is known in physics as the \textit{microcanonical} measure.

\item[b.)] In a system at \textit{constant temperature} $T$ particles exchange energy
with the walls of the box $\Omega$ in such a way that the mean energy is fixed.
The appropriate macrovariables are then $X = (T, V, N)$ and the equilibrium state
is of the form
$$
P_{T,V,N} \sim \exp[-\beta H], \;\;\; \beta = 1/T,
$$
known as the \textit{canonical} measure.

\end{itemize}
Having roughly characterized 
equilibrium states, we may say that \textit{nonequilibrium}
states arise whenever the conditions for the establishment of equilibrium
are not fulfilled. As such, this definition is about as useful as it would
be to define some area of biology as the study of non-elephants. We can be 
somewhat more precise by making a distinction between 
\begin{itemize}

\item[(i)] \textit{Systems approaching equilibrium.} By definition, 
the macrostate of such a system is time-dependent. In addition, systems in
this class often become spatially inhomogeneous; an important and much studied
case are systems undergoing phase separation \cite{Bray1994}. 

\item[(ii)] \textit{Nonequilibrium stationary states (NESS).} These
systems are kept out of equilibrium by external influences. They
are stationary, in the sense that macroscopic state variables are
time-independent, and they may or may not be spatially homogeneous.
In any case they are characterized by non-vanishing macroscopic currents.

\end{itemize}
  
Examples for NESS are 

\begin{itemize}

\item \textit{Heat conduction.} In a system with boundaries
held at different temperatures there is a stationary
energy current proportional to the temperature gradient
(\textit{Fourier's law}). 

\item \textit{Diffusion.} In a system coupled to particle reservoirs
held at different densities there is a mass current proportional
to the density gradient (\textit{Fick's law}).

\item \textit{Electric conduction.} Here particles are charged and
move under the influence of a constant electric field. The particle
current is proportional to the field strength (\textit{Ohm's law}).

\end{itemize}
Among these three examples, 
the first two can be further characterized as \textit{boundary driven},
in the sense that the NESS is maintained by boundary conditions 
on the quantity that is being transported (heat, mass),
whereas the last example illustrates a \textit{bulk-driven} NESS
maintained by an external field acting in the bulk of the system.

NESS are the simplest examples of nonequilibrium states.
Nevertheless, their description in the framework of classical Hamiltonian
mechanics is conceptually subtle and technically demanding (see, e.g., \cite{Vollmer2002}).
The main reason is that a Hamiltonian system under constant driving inevitably 
accumulates energy. In order to allow for the establishment of a steady state,
dissipation has to be introduced through the coupling to an external 
reservoir, that is, a system with an infinite number of degrees of freedom.

These difficulties can be avoided by starting from \textit{stochastic}
microscopic dynamics. While less realistic on the microscopic level,
stochastic models provide a versatile framework for addressing
fundamental questions associated with the behavior of many-particle
systems far from equilibrium. The class of models of interest here are known
in the probabilistic community as \textit{interacting particle systems}.
These are lattice models with a discrete (finite or infinite) set of states
associated with each lattice site and local interactions. We focus specifically
on exclusion processes, which are introduced in the next section.

It is worth pointing out that 
the notion of equilibrium states in statistical physics, as outlined above, 
is much more restrictive than the usage of the corresponding term in most areas of mathematics, where
an \textit{equilibrium} is commonly understood to be any time-independent solution
of some deterministic or stochastic time evolution. Thus NESS are equilibria in 
the mathematical sense. In Sect.~\ref{Markov} we will give a precise definition of 
what distinguishes physical equilibria from other time-independent
states in the context of continuous time Markov chains.

\section{An introduction to exclusion processes}
\label{Exclusion}

\subsection{Definition}
\label{jsec3.1}

The simple exclusion process was introduced in 1970 by Frank Spitzer
\cite{Spitzer1970}. Particles occupy the sites of a 
$d$-dimensional lattice, which for the purposes of this discussion will be
taken to be a finite subset $\Omega \subset \mathbb{Z}^d$. The particles
are indistinguishable, which implies that a microstate or configuration
of the system is given by 
$$
\eta = \{ \eta_x \}_{x \in \Omega} \in \{ 0,1 \}^\Omega,
$$
where $\eta_x = 0$ (1) if site $x$ is vacant (occupied). The dynamics
can be informally described as follows (for a detailed construction
see \cite{Spitzer1970,Liggett1999}):

\begin{itemize}

\item Each particle carries a clock which rings according to a Poisson
process with unit rate (i.e., the waiting times between rings are
exponentially distributed).

\item When the clock rings the particle at site $x$ selects a 
target site $y$ with probability $q_{xy}(\eta)$ and attempts to jump there.

\item The jump is performed if the target site is vacant and discarded
otherwise; this step implements the \textit{exclusion interaction}
between particles and enforces the single occupancy constraint
$\eta_x = 0$ or 1.

\end{itemize}
Together these rules define the exclusion process as a continuous time Markov
chain on a finite state space; some general properties of such chains will
be discussed in the next section. \textit{Interactions} (beyond the exclusion
interaction) can be introduced through the dependence of the jump matrix 
$q_{xy}$ on the configuration $\eta$. Similarly,
\textit{inhomogeneity} associated with sites or particles can be introduced
by letting the waiting times and the jump matrix depend
explicitly on the particle positions or the particle labels, 
see \cite{Krug2000}.

We next restrict the discussion to the one-dimensional case
with nearest neighbor hopping and without inhomogeneities or explicit
interactions. Then
$$
q_{xy} = q \delta_{y,x+1} + (1-q) \delta_{y,x-1}.
$$
Informally, the particle attempts to jump to the right with
probability $q$ and to the left with probability $1-q$. 
The following cases are of interest:
\begin{itemize}

\item[(i)] $q = 1/2$ defines the \textit{symmetric 
simple exclusion process}
(SSEP). We will see below that this is really an equilibrium system.
However, when defined on a finite lattice of sites $x = 1,...,L$ 
and supplemented with boundary rates $\alpha, \beta, \gamma, \delta$
which govern the injection $(\alpha, \delta)$ and extraction 
$(\gamma, \beta)$ of particles at the boundary sites $i = 1$ and 
$i = L$, this model provides a nontrivial example for a boundary-driven
NESS \cite{Derrida2007}. 

\item[(ii)] $q \neq 1/2$ defines the \textit{asymmetric 
simple exclusion process} (ASEP). When considered on the one-dimensional
ring (a lattice with \textit{periodic boundary conditions}) the
system attains a bulk-driven NESS in which there is a non-vanishing
stationary mass current. This is the simplest realization
of a \textit{driven diffusive system} \cite{Schmittmann1995}.

Note that the boundary conditions
are crucial here. On a finite lattice with closed ends, which prevent particles
from entering or leaving the system, an \textit{equilibrium} state is established
in which the bias in the jump probability is compensated by a density gradient;
this is the discrete analog of a gas in a gravitational field, as 
described by the barometric formula. Another possibility is to consider
a finite lattice with open ends at which particles are injected and extracted
at specified rates \cite{Krug1991}. This leads to a NESS with a surprisingly
complex structure, see \cite{Blythe2007} for review.

\item[(iii)] $q = 1$ (or 0) defines the \textit{totally asymmetric 
simple exclusion process} (TASEP). 
In contrast to the case of general $q$,
this process can also be formulated in discrete time \cite{Yaguchi1986}: 
In one time step $t \to t+1$, all particles attempt to move to the right
(say) simultaneously and independently with probability $\pi \in (0,1]$;
moves to vacant sites are accepted and moves to occupied sites discarded.
Such a discrete time dynamics cannot be defined for $0 < q < 1$, because it would
lead to conflicts when different particles attempt to simultaneously 
access the same vacant site.

For $\pi \to 0$ the discrete time TASEP (\textit{dTASEP}) reduces 
to the continuous time process in
rescaled time $\pi t$, while for $\pi = 1$ it becomes a deterministic
cellular automaton which has number 184 in Wolfram's classification
\cite{Wolfram1983,Krug1988}. 
The case of general $\pi$ has been studied mostly in the context of
vehicular traffic modeling \cite{Schreckenberg1995,Chowdhury2000}. 

Note that in terms
of the waiting time picture sketched above, the discrete time dynamics
corresponds to replacing the exponential waiting time distribution 
by a geometric distribution with support on integer times only.
The exponential and geometric waiting time distributions are the only ones
that encode \textit{Markovian} dynamics \cite{Krug1998}.
The waiting time representation will play an important role in the exact solution
of the dTASEP presented below in Sect.\ref{tsec:C}.

\end{itemize}

\subsection{Continuous time Markov chains}
\label{Markov}

Before discussing some specific properties of exclusion processes, we 
outline the general setting of continuous time Markov chains
(see \cite{Resnick2002} for an introduction). Consider a Markov chain
with a finite number of states $i=1,\ldots,C$ and transition rates
$\Gamma_{ij}$. The rates define the dynamics in the following way:

\begin{itemize}

\item[]
When the chain is in state $i$ at time $t$, a transition to state $j \neq i$
occurs in the time interval $[t,t+dt]$ with probability $\Gamma_{ij} dt$.

\end{itemize}
The key quantity of interest is the transition probability
$$
P_{ki}(t) = \textrm{Prob}[\textrm{state} \; i \; \textrm{at} \; t \vert
\textrm{state} \; k \; \textrm{at} \; 0] \equiv P_i(t)
$$
where the initial state $k$ is included through the initial condition
$P_i(0) = \delta_{ik}$. The transition probability satisfies the 
evolution equation 
\begin{equation}
\label{master}
\frac{d}{dt} P_i = \sum_{j \neq i} \Gamma_{ji} P_j - \sum_{j \neq i} \Gamma_{ij} P_i 
= \sum_j A_{ji} P_j,
\end{equation}
which is known as the \textit{master equation} in physics \cite{vanKampen2001}
and as the \textit{forward equation} in the theory of stochastic processes \cite{Resnick2002}.
Here the \textit{generator matrix}
$$
A_{ij} = \left\{ \begin{array}{l@{\quad:\quad}l}
\Gamma_{ij}  & i \neq j  \\ 
-\sum_{k \neq i} \Gamma_{ik} & i = j \end{array} \right.
$$
has been introduced. The master equation simply accounts for the balance
of probability currents going in and out of each state of the Markov chain.
To bring out this structure we rewrite (\ref{master}) in the form 
\begin{equation}
\label{cont}
\frac{d}{dt} P_i = \sum_j K_{ij}, \;\;\;\; 
K_{ij} = \Gamma_{ji} P_j - \Gamma_{ij} P_i,
\end{equation}
where $K_{ij}$ is the \textit{net probability current} between states $i$ and 
$j$ \cite{Zia2007}. If the chain is \textit{irreducible}, in the sense that
every state can be reached from every other state through a connected
path of nonzero transition rates, the solution of (\ref{master}) approaches
at long times a unique, stationary invariant measure $P_i^\ast$ determined
by the condition
\begin{equation}
\label{eigen}
\sum_j A_{ji} P_j^\ast = 0.
\end{equation}
The invariant measure is the left eigenvector of the generator matrix, with
eigenvalue zero. Based on (\ref{cont}) we can rewrite (\ref{eigen}) as
\begin{equation}
\label{currents}
\sum_j K^\ast_{ji} = 0 \;\;\; 
\textrm{with} \;\;\; K_{ji}^\ast = \Gamma_{ji} P_j^\ast - \Gamma_{ij} P_i^\ast.
\end{equation}
Two classes of Markov chains may now be distinguished depending on how the
stationarity condition (\ref{currents}) is realized:

\begin{itemize}

\item[(i)] $K_{ij}^\ast = 0 \; \forall \; i,j$. In this case the
probability currents cancel between any two states $i,j$, 
\begin{equation}
\label{detbal}
\Gamma_{ij} P_i^\ast = \Gamma_{ji} P_j^\ast,
\end{equation}
a condition that is known in physics as \textit{detailed balance}. 
In the mathematical literature Markov chains with this property are 
called \textit{reversible}, because (\ref{detbal}) implies that the
weight of any trajectory (with respect to the invariant measure) 
is equal to that of its image under time-reversal
\cite{Resnick2002,Kelly1979}. Detailed balance or, equivalently, reversibility
is a fundamental property that any stochastic model of a physical
system \textit{in equilibrium} must satisfy, because equilibrium
states are distinguished by invariance under time reversal
% \footnote{This statement
% has to be somewhat modified in the presence of magnetic fields.}. 

\item[(ii)] $K_{ij}^\ast \neq 0$ at least for some pairs of states
$i,j$. Such a Markov chain is irreversible and describes a system in a
NESS.

\end{itemize} 
Examples for both kinds of situations will be encountered in the next section.

\subsection{Stationary measure of the exclusion process}
\label{Measures}

We consider the ASEP on a ring of $L$ sites 
with a fixed number $N$ of particles. The total number of microstates 
$\eta$ is then $C = {L \choose N}$ and the transition
rates are
\begin{equation}
\label{ASEPrates} 
\Gamma(\eta \to \eta') = \left\{ \begin{array}{l@{\quad:\quad}l}
q  & (...\bullet \circ ...) \to (...\circ \bullet ...) \\ 
1-q  & (...\circ \bullet ...) \to (...\bullet \circ ...) \\
0 & \textrm{else}. \end{array} \right.
\end{equation}
Here $(...\bullet \circ ...)$ denotes a local configuration
with an occupied site $(\bullet)$ to the left of a vacant site
$(\circ)$, and it is understood that only configurations 
$\eta, \eta'$ that differ by the exchange of a single particle-vacancy
pair are connected through nonzero transition rates. The stationary
measure $P^\ast(\eta)$ is determined by the condition 
\begin{equation}
\label{statASEP}
\sum_{\eta'} \Gamma(\eta' \to \eta) P^\ast(\eta') = 
\sum_{\eta'} \Gamma(\eta \to \eta') P^\ast(\eta) 
\;\;\;\; \forall \; \eta.
\end{equation}
As the simplest possibility, let us assume that the invariant measure is 
uniform on the state space,
\begin{equation}
\label{uni}
P^\ast(\eta) = {L \choose N}^{-1} \;\;\;
\Rightarrow \;\;\;
K^\ast(\eta, \eta') = [\Gamma(\eta' \to \eta) - 
\Gamma(\eta \to \eta') ] {L \choose N}^{-1}.
\end{equation}
We discuss separately the symmetric and the asymmetric process.

\begin{itemize}

\item $q=1/2$ (SSEP). Here the rate $q = 1-q = 1/2$ for all allowed processes, and for each 
allowed process the reverse process occurs at the same rate. We conclude that detailed
balance holds in this case, $K^\ast = 0$, and the SSEP is reversible as announced previously.

\item $q \neq 1/2$ (ASEP). Because for any allowed process with rate $q$ the reverse process
occurs at rate $1 - q \neq q$ and vice versa, 
detailed balance is manifestly broken, $K^\ast \neq 0$,
and we are dealing with an irreversible NESS. However, we now show that the uniform 
measure (\ref{uni}) is nevertheless invariant. To see this, consider the 
total transition rates for all processes leading into or out of a given configuration
$\eta$. We have 
$$
\Gamma_{\mathrm{tot}}^\mathrm{in}(\eta) = \sum_{\eta'} \Gamma(\eta' \to \eta) = 
q {\cal{N}}_{\circ \bullet}(\eta) + (1 - q) {\cal{N}}_{\bullet \circ}(\eta)
$$
where ${\cal{N}}_{\circ \bullet}(\eta)$ denotes the number of pairs of sites with
a particle to the right of a vacancy in the configuration $\eta$. Similarly
$$
\Gamma_{\mathrm{tot}}^\mathrm{out}(\eta) = \sum_{\eta'} \Gamma(\eta \to \eta') = 
q {\cal{N}}_{\bullet \circ}(\eta) + (1 - q) {\cal{N}}_{\circ \bullet}(\eta).
$$
A little thought reveals that ${\cal{N}}_{\bullet \circ}(\eta) = {\cal{N}}_{\circ \bullet}(\eta)$
for any configuration $\eta$. Hence $\Gamma_{\mathrm{tot}}^\mathrm{in}(\eta) = 
\Gamma_{\mathrm{tot}}^\mathrm{out}(\eta)$ for any $q$, and the stationarity condition
(\ref{statASEP}) is satisfied for the uniform measure (\ref{uni}). 
 
\end{itemize}
A few remarks are in order.

\begin{itemize}

\item[(i)] The invariance of the uniform measure (\ref{uni}),
and the fact that it is independent of the bias $q$, relies
crucially on the ring geometry. With open boundary conditions allowing for the injection and extraction
of particles both the SSEP and the ASEP display nontrivial invariant measures 
characterized by long-ranged correlations and the possibility of 
phase transitions \cite{Derrida2007,Blythe2007}. For example, for the SSEP with
boundary densities $\rho_L$ at $x=1$ and $\rho_R$ at $x=L$ one finds a linear
mean density profile, as expected from Fick's law, but in addition there are 
long-ranged density-density correlations on the scale $L$,
which take the form \cite{Derrida2007,Spohn1983}
$$
\mathbb{E} (\eta_{L\xi} \eta_{L \xi'}) -  
\mathbb{E} (\eta_{L\xi}) \mathbb{E}(\eta_{L \xi'}) =
-\frac{\xi (1 - \xi')}{L} (\rho_L - \rho_R)^2.
$$
Here $\xi, \xi^\prime \in [0,1]$ are scaled position variables with
$\xi < \xi^\prime$.

\item[(ii)] The invariant measure of the dTASEP on the ring is 
\textit{not} uniform. Rather, one finds
a Gibbs measure with repulsive nearest-neighbor interactions between
the particles \cite{Yaguchi1986,Schadschneider1993,Schreckenberg1995}. 
This means that the probability of a configuration can be written as
a product of pair probabilities,
\begin{equation}
\label{Ising}
P_\rho^\ast(\eta) \sim \prod_{x} p_\rho(\eta_x,\eta_{x+1}),
\end{equation}
where the limit $N, L \to \infty$ at fixed density 
\begin{equation}
\label{density}
\rho = \mathbb{E}(\eta_x) = N/L
\end{equation}
is implied and 
\begin{equation}
\label{DTASEP1}
p_\rho(0,1) = p_\rho(1,0) = \frac{1 - \sqrt{1 - 4 \pi \rho(1- \rho)}}{2\pi},  
\end{equation}
\begin{equation}
\label{DTASEP2}
p_\rho(0,0) = 1 - \rho - p_\rho(1,0), \;\; 
p_\rho(1,1) = \rho - p_\rho(1,0). 
\end{equation}
For $\pi \to 0$ this reduces to a Bernoulli measure of independent particles
(see Sect.~\ref{Hydro}), whereas for $\pi \to 1$ we have $p_\rho(1,0) \to (1 - \vert 1 - 
2 \rho \vert)/2$, which implies that $p_\rho(1,1) \to 0$ for $\rho < 1/2$
and $p_\rho(0,0) \to 0$ for $\rho > 1/2$. At $\pi = 1$ and mean density $\rho = 1/2$ the 
measure is concentrated on the two configurations $\eta_x = [1 \pm (-1)^{x}]/2$.  

\item[(iii)] The invariance of the uniform measure for the ASEP on the ring
is an example of \textit{pairwise balance}
\cite{Schuetz1996}, a property that generalizes the detailed
balance condition (\ref{detbal}) into the form
$$
\Gamma(\eta \to \eta') P^\ast(\eta) = \Gamma(\eta'' \to \eta) P^\ast(\eta'').
$$
This means that for each configuration $\eta'$ contributing to the outflux of
probability out of the state $\eta$ there is a configuration $\eta''$ whose influx
contribution precisely cancels that outflux. In other words, the terms in the sums
on the two sides of (\ref{statASEP}) cancel pairwise.

\end{itemize}

\subsection{Hydrodynamics}
\label{Hydro}

An important goal in the study of stochastic interacting particle systems is to understand
how deterministic evolution equations emerge from the stochastic microscopic dynamics
on large scales \cite{Spohn1991,Kipnis1999,Bertini2007}. 
This is similar to the (much harder) problem
of deriving hydrodynamic equations from the Newtonian dynamics of molecules in a gas
or a fluid. The mathematical procedure involved in the derivation of macroscopic
evolution equations for systems with conserved quantitities is therefore referred
to as the \textit{hydrodynamic limit}. Here we give a heuristic sketch of hydrodynamics
for the ASEP. 

The key input going into the hydrodynamic theory is the relationship between the particle
density $\rho$ and the stationary particle current $J$. 
The particle current is defined as the net number
of particles jumping from a site $x$ to the neighboring site 
$x+1$ per unit time, which is independent of $x$ in the stationary state.
From the definition of the ASEP we have 
$$
J = q \mathbb{E}[\eta_x (1 - \eta_{x+1})] - (1-q) \mathbb{E}[\eta_{x+1} (1 - \eta_{x})]
$$
where expectations are taken with respect to the invariant
measure. Since all configurations of $N$ particles on the lattice of $L$ sites
are equally probable, 
$$
\mathbb{E}[\eta_x (1 - \eta_{x+1})]  = \mathbb{E}[\eta_{x+1} (1 - \eta_{x})]
 = \frac{N}{L} \frac{(L-N)}{L-1}.
$$ 
This is just the probability of finding a filled site next to a vacant site, 
which is obtained by first placing one out of $N$ particles in one of $L$ sites,
and then placing one out of $L-N$ vacancies in one of 
the remaining $L-1$ sites. We conclude
that 
$$
J = \frac{(2 q -1) \rho(1- \rho)}{1 - 1/L} \to (2 q -1) \rho(1 - \rho) \;\;
\textrm{for} \;\; L \to \infty,
$$ 
where the particle density (\ref{density}) is kept fixed. Similarly 
\begin{equation}
\label{Bernouilli}
\mathbb{E}[\eta_x \eta_y ] = \frac{N}{L}\frac{N-1}{L-1} \to \rho^2 =
\mathbb{E}[\eta_x] \mathbb{E}[\eta_y]  
\;\; \textrm{for} \;\; L \to \infty
\end{equation}
for any pair of sites $x \neq y$. This implies that in the invariant measure
on the ring, restricted to a fixed finite number of sites,
for $L \to \infty$ each site is occupied independently with
probability $\rho$ (Bernoulli measure).

We can now formulate the basic idea of the hydrodynamic limit \cite{Lebowitz88}. 
Suppose that we start the ASEP at time $t = 0$ from a Bernoulli measure
with a slowly varying density $\rho(x,0)$. Here ``slowly varying'' means
that variations occur on a scale $\ell \gg 1$. Since the invariant measure
of the ASEP is a Bernoulli measure of \textit{constant} density,
it is plausible that, if $\ell$ is chosen large enough, the evolving measure
will remain close to a Bernoulli measure with a time and space dependent
density $\rho(x,t)$ at all times; and because the particle density is locally
conserved, the evolution equation for $\rho(x,t)$ must be of conservation
type,
\begin{equation}
\label{hydro1}
\frac{\partial}{\partial t} \rho(x,t) + \frac{\partial}{\partial x}
j(x,t) = 0.
\end{equation}
In the limit $\ell \to \infty$ we may expect, in the spirit of a law of large
numbers, that the local particle current $j(x,t)$ converges to the 
stationary current associated with the local density $\rho(x,t)$,
$$
j(x,t) \to J(\rho(x,t)),
$$
such that (\ref{hydro1}) becomes an autonomous, deterministic hyperbolic conservation law
\begin{equation}
\label{hydro2}
\frac{\partial \rho}{\partial t} + \frac{\partial}{\partial x} J(\rho) = 0
\end{equation}
for the density profile $\rho(x,t)$. Equation (\ref{hydro2}) is known as the 
\textit{Euler} equation for the ASEP, because similarly to the Euler equation
in fluid mechanics it lacks a second order ``viscosity'' term $\nu
\partial^2 \rho/\partial x^2$. It must be emphasized that such a term does
\textit{not} appear to leading order, when the hydrodynamic limit is carried out
at fixed $q \neq 1/2$. It is only present in the \textit{weakly asymmetric}
case, which implies that $q \to 1/2$ in the limit $\ell \to \infty$
such that $\ell (q - 1/2)$ is kept fixed \cite{DeMasi1989}.

The Euler equation (\ref{hydro2}) has been rigorously established for a wide
range of models, including cases in which the invariant measure and the
current-density relation $J(\rho)$ are not explicitly known \cite{Seppalainen1999}. 
We conclude this section by a brief discussion of the properties of the nonlinear
PDE (\ref{hydro2}), assuming a general (but convex) current-density relation
with $J(0) = J(1) = 0$. This includes in particular the dTASEP for 
which 
\begin{equation}
\label{JTASEP}
J(\rho) = \pi p_\rho(1,0) = \frac{1}{2}[1 - \sqrt{1 - 4 \pi \rho(1 -\rho)}].
\end{equation}

\begin{itemize}

\item[(i)] \textit{Shock formation}. Hyperbolic conservation laws
of the form (\ref{hydro2}) can generally be solved by the 
\textit{method of characteristics}. To this end we first 
rewrite (\ref{hydro2}) in the form
\begin{equation}
\label{cEq}
\frac{\partial \rho}{\partial t} + c(\rho) \frac{\partial \rho}{\partial x} = 0,
\end{equation}
where 
\begin{equation}
\label{speed}
c(\rho) = \frac{dJ}{d\rho}.
\end{equation}
A characteristic is a trajectory of a point of constant density, and the key observation
is that the characteristics of (\ref{cEq}) are straight lines. Denoting by $v_{\rho_0}(x,t)$ the 
position of a point of density $\rho_0 = \rho(x,0)$ at time $t$, we have to 
satisfy the condition
$$
\rho(v_{\rho_0}(x,t),t) = \rho_0 = \rho(x,0)
$$
at all times. Taking the time derivative of this relation and using (\ref{cEq}) we see that
the solution is
$$
v_{\rho_0}(x,t) = x + c(\rho_0)t,
$$
i.e. points of constant density travel at the \textit{kinematic wave speed} (\ref{speed}). 

The convexity of the current-density relation implies that $c(\rho)$ is a 
decreasing function of the density. As a consequence characteristics collide
in regions of increasing initial density, $d \rho(x,0)/dx > 0$, leading
to the formation of density discontinuities (\textit{shocks}) in finite time.
At this point the description by the PDE (\ref{hydro2}) breaks down,
but the speed $V$ of a shock separating regions of density $\rho_L$
on the left and $\rho_R > \rho_L$ on the right is easily inferred from
mass conservation to be given by 
\begin{equation}
\label{shockspeed}
V = \frac{J(\rho_R)-J(\rho_L)}{\rho_R - \rho_L}.
\end{equation}
Note that $V \to c$ for $\rho_L \to \rho_R$.
On the microscopic level shocks are represented by 
the \textit{shock measures} of the ASEP \cite{Ferrari1991,Ferrari1992}. 
These are inhomogeneous invariant
measures on $\mathbb{Z}$ which approach Bernoulli measures with density
$\rho_L$ and $\rho_R$ for $x \to -\infty$ and $x \to \infty$, respectively.
The microscopic structure of shocks has been studied in considerable
detail \cite{Derrida1993}.

\item[(ii)] \textit{Rarefaction waves.} If the initial density profile is a step
function
\begin{equation}
\label{step}
\rho(x,0) = \left\{ \begin{array}{l@{\quad:\quad}l}
\rho_L  & x < 0 \\ 
\rho_R & x > 0 \end{array} \right.
\end{equation}
with $\rho_L > \rho_R$, a diverging fan of characteristics forms leading
to a broadening, self-similar density profile
\begin{equation}
\label{wave}
\rho(x,t) = \left\{ \begin{array}{l@{\quad:\quad}l}
\rho_L  & x < c(\rho_L) t \\
\rho_R & x > c(\rho_R) t \\ 
\phi(x/t) & c(\rho_L) < x/t < c(\rho_R), \end{array} \right.
\end{equation}
where the shape function $\phi(\xi)$ can be computed from the current-density
relation $J(\rho)$. Inserting the ansatz (\ref{wave}) into
(\ref{cEq}) we see that 
\begin{equation}
\label{phic}
\phi(\xi) = c^{-1}(\xi).
\end{equation}
For the continuous time ASEP the interpolating shape is linear, 
$$
\phi(\xi) = \frac{1}{2} \left(1 - \frac{\xi}{2q-1} \right).
$$

\end{itemize}
 
\subsection{Mapping to a growth model}
\label{Growthmodel}

\begin{figure}
\begin{center}
\resizebox{12cm}{!}{\includegraphics{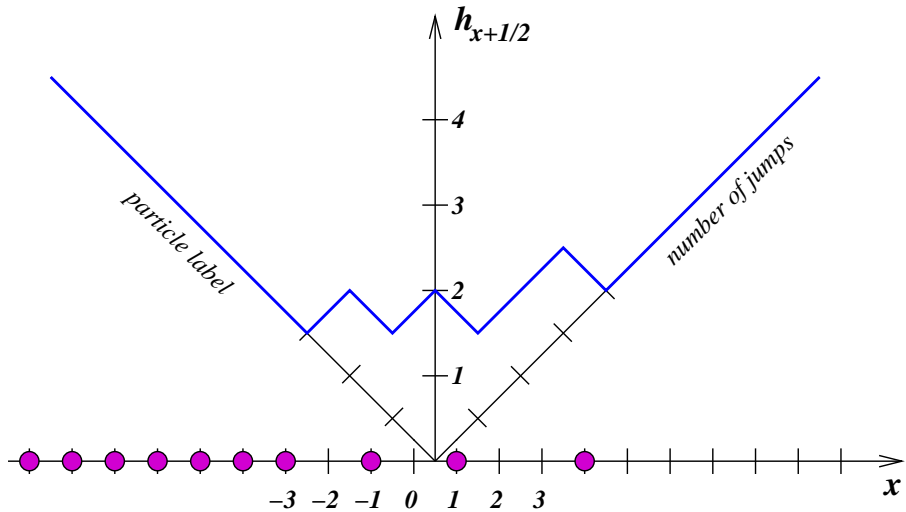}}
\end{center}
\caption{(Color online) Schematic of the mapping between a configuration of the ASEP and the corresponding height
configuration $h_{x+\frac{1}{2}}$ (bold blue line). Initially all sites $x \leq 0$ in the ASEP
are occupied, and all sites $x > 0$ are vacant (step initial condition). The ASEP occupation variables
determine the height differences according to (\ref{ASEPslope}). At the same time, the height 
increment $h_{x+\frac{1}{2}}(t) - h_{x+\frac{1}{2}}(0)$ counts the net number of particles that have 
crossed the bond $(x,x+1)$ from left to right up to time $t$. In the figure two particles 
have crossed the origin, and $h_{\frac{1}{2}} = 2$. Rotating the height configuration
by 45$^\circ$ provides a representation of the net number of jumps a given particle has undergone
as a function of the particle label, counted backwards from $x=0$ (the $l$'th particle
is the one that was located at $x = - l$ at time $t=0$).}
\label{Rostfigure}
\end{figure}

The representation of the one-dimensional ASEP as a growth model
seems to have been formulated first by Rost \cite{Rost1981}.
It is illustrated in Fig.~\ref{Rostfigure} for a step initial condition,
where all sites to the left of the origin are occupied ($\eta_x = 1$ for
$x \leq 0$) and all sites to the right of the origin are empty ($\eta_x = 0$
for $x > 0$). This initial condition will also play a central role below in Sect.\ref{tsec:R}. 

The mapping assigns to every configuration $\eta = \{\eta_x \}$ of the ASEP
a configuration of height variable $\{h_{x+\frac{1}{2}} \}$,
where the shift of the index indicates that the height variable $h_{x+\frac{1}{2}}$
lives on the bond connecting the ASEP sites $x$ and $x+1$. After fixing the height at
a reference point, e.g. by setting $h_{\frac{1}{2}}$ = 0, the height configuration
is uniquely determined by the relation
\begin{equation}
\label{ASEPslope}
h_{x+\frac{1}{2}} - h_{x-\frac{1}{2}} = \frac{1}{2} - \eta_x.
\end{equation}    
The ASEP occupation variable encode the local slopes of the height profile,
which take the values $h_{x+\frac{1}{2}} - h_{x-\frac{1}{2}} = \pm \frac{1}{2}$, hence the
name ``single step model'' in the physics literature \cite{Meakin86,Plischke87}. 

The step initial condition corresponds to the initial height profile 
$$h_{x+\frac{1}{2}}(0) = \frac{1}{2} \vert x \vert,$$
which accounts for the designation as a ``corner growth model'' in the
mathematical literature.
It can be seen from Fig.~\ref{Rostfigure} that a particle jumping 
across a bond to the 
right (left) increases (decreases) the corresponding height variable by
one unit. Thus the height $h_{x+\frac{1}{2}}$ is an odd (even) multiple of 
$\frac{1}{2}$ for odd (even) values of $x$, and the height change 
$h_{x+\frac{1}{2}}(t) - h_{x+\frac{1}{2}}(0)$ is equal to the net number of particles
that have jumped across the bond $(x, x+1)$ from left to right up to time 
$t$ (jumps from right to left are counted with a negative sign). Finally, 
for the special case of 
the step initial condition, the net number of jumps (forward minus
backward) performed by a given particle
can also be read off from the height profile (see Fig.~\ref{Rostfigure}). 

The mapping (\ref{ASEPslope}) is clearly not restricted to the step initial 
condition. Of particular interest are translationally invariant initial 
conditions, which can be constructed deterministically or stochastically.
For example, to generate a deterministic initial condition of density $\rho = 1/n$, 
one simply places a particle at every $n$'th site of the lattice, and a 
stochastic initial condition is obtained by occupying sites independently
with probability $\rho$. The two types
of initial conditions differ in the \textit{roughness} of the corresponding
height configuration, which is quantified by the \textit{height
difference correlation function}
\begin{equation}
\label{hdiff}
G(r) = \mathbb{E}[(h_{y+r} - h_{y})^2] - \mathbb{E}[h_{y+r} - h_y]^2.
\end{equation}
An ensemble of height configurations on $\mathbb{Z}$ is said to be \textit{smooth} if 
$\lim_{r \to \infty} G(r) < \infty$ and \textit{rough} otherwise \cite{Krug1991b}. The deterministic
initial conditions described above are smooth in this sense, whereas for the 
stochastic initial condition a simple computation using (\ref{ASEPslope}) and 
(\ref{Bernouilli}) shows that
\begin{equation}
\label{ASEPrough}
G(r) = \rho (1 - \rho) \vert r \vert.
\end{equation}

\section{The KPZ conjecture}

\label{KPZ_conjecture}

The asymmetric exclusion process and the equivalent growth model
introduced in the preceding subsection are representatives of a large class of
models, which was brought to the forefront
of research in nonequilibrium statistical physics in 1986 by a 
seminal paper of Kardar, Parisi and Zhang (KPZ) \cite{Kardar1986}. Working in the
framework of a phenomenological stochastic continuum description, they formulated
what may be called a \textit{universality hypothesis} encompassing the fluctuation
properties of a large class of different microscopic models\footnote{For an introduction
to the idea of universality from a mathematical perspective see \cite{TDeift07}.}. The classic period 
of research in this area has been extensively reviewed in the literature 
\cite{Krug1991b,HalpinHealy1995,Krug1997}. Here we
aim to give a concise and simple presentation of the KPZ conjecture, in order
to place the more recent developments (to be elaborated in the following sections)
into their proper context. 

\subsection{The Kardar-Parisi-Zhang equation}

We start from the hydrodynamic equation (\ref{hydro2}) with a general current-density
relation $J(\rho)$. Since we are interested in fluctuations around a
state of constant mean density $\bar \rho$, we write $\rho(x,t) = \bar \rho + u(x,t)$
and expand to second order in $u$, which yields
\begin{equation}
\label{expansion}
\frac{\partial u}{\partial t} = - c(\bar \rho) \frac{\partial u}{\partial x}
- \lambda u \frac{\partial u}{\partial x},
\end{equation}
where 
\begin{equation}
\label{lambda}
\lambda = \frac{d^2J}{d^2\rho}(\bar \rho).
\end{equation} 
The linear drift term on the right hand side
can be eliminated by a Galilei transformation $x \to x - ct$, which leaves us with
what is known (for $\lambda = 1$) as the \textit{inviscid Burger equation}. 

Now fluctuations
are introduced (in the spirit of fluctuating hydrodynamics \cite{Spohn1991})  
by adding a random force to the right hand side of (\ref{expansion}). In order to
guarantee mass conservation, this term must take the form of a derivative
$-\partial \zeta/\partial x$ of a stochastic process $\zeta(x,t)$ in space and time.
This is assumed to be a stationary Gaussian process with zero mean and a covariance function
\begin{equation}
\label{covariance}
\mathbb{E}[\zeta(x,t) \zeta(x',t')] = a_x^{-1} a_t^{-1} G[(x - x')/a_x, (t - t')/a_t]
\end{equation}
which vanishes beyond a small correlation length $a_x$ and a short correlation time $a_t$.
Usually one takes formally\footnote{In the hydrodynamic context \cite{Forster1977}
it is also of interest to consider the solutions of (\ref{noisy}) on scales 
\textit{small} compared to the spatial driving scale $a_x$.} $a_x, a_t \to 0$, 
which reduces the right hand side of (\ref{covariance})
to a product of $\delta$-functions,
\begin{equation}
\label{white}
\mathbb{E}[\zeta(x,t) \zeta(x',t')] \to D \delta(x - x') \delta(t - t')
\end{equation}
and turns the process $\zeta(x,t)$ into spatio-temporal white noise of strength $D$.
This rather violent driving has to be compensated by a small viscosity term 
$\nu \partial^2 u/ \partial x^2$ with $\nu > 0$. Putting all ingredients together we thus arrive at
the \textit{stochastic Burgers equation}
\begin{equation}
\label{noisy}
\frac{\partial u}{\partial t} = \nu \frac{\partial^2 u}{\partial x^2} 
- \lambda u \frac{\partial u}{\partial x} + \frac{\partial \zeta}{\partial x}
\equiv - \frac{\partial}{\partial x} j(x,t),
\end{equation}
first introduced in the context of randomly stirred
fluids \cite{Forster1977} and subsequently applied to fluctuations in 
the exclusion process by van Beijeren, Kutner and Spohn \cite{vanBeijeren1985}.
% Here $j(x,t)$ stands for the local current fluctuation (recall that the 
% macroscopic mean current has been subtracted). 

To establish the connection to growth models we proceed in analogy to
the discrete case discussed in Sect.~\ref{Growthmodel}. We introduce the 
\textit{height function} $h(x,t)$ through the time-integrated 
particle current,
\begin{equation}
\label{height}
h(x,t) = \int_0^t j(x,s) ds,
\end{equation}
Supplementing this with the initial condition $u(x,0) = 0$, it follows
from the conservation law for $u$ that 
\begin{equation}
\label{hu}
\frac{\partial h}{\partial x} = -u, 
\end{equation}
and therefore
\begin{equation}
\label{KPZ}
\frac{\partial h}{\partial t} = \nu \frac{\partial^2 h}{\partial x^2} +
\frac{\lambda}{2} \left( \frac{\partial h}{\partial x} \right)^2 - \zeta,
\end{equation}
which is precisely the KPZ-equation \cite{Kardar1986}. In general there is also
a constant term on the right hand side of (\ref{KPZ}) which has been set to zero.
The defining feature of the equation is the quadratic nonlinearity on the right hand side, which is present whenever 
$\lambda \neq 0$, that is, when the current is a (generic) nonlinear
function of the density\footnote{Note that it is possible to have
  $\lambda \neq 0$ even if the current itself vanishes at the specific
  mean density under consideration (see \cite{TBoFe08} for an example).} 
[compare to (\ref{lambda})]. 

It is important to clearly understand the relation between
the stochastic PDE's (\ref{noisy},\ref{KPZ}) and the underlying discrete particle systems. 
The coefficient $\lambda$ in (\ref{KPZ}) is defined through the current-density
relation according to (\ref{lambda}), but the viscosity $\nu$ and the noise strength $D$
in (\ref{white}) do not directly appear on the discrete level. To give these coefficients
a consistent interpretation, we start from the observation \cite{Forster1977,Huse1985} that
the invariant measure of (\ref{noisy}) with spatio-temporal 
white-noise driving is spatial white noise\footnote{This remains true
  for certain discretizations of (\ref{KPZ}) \cite{Krug1991b,TSaSp09}.} with
strength $D/2 \nu$. This is easy to check for the linearized equation ($\lambda = 0$) but
it remains true also for $\lambda \neq 0$, somewhat analogous to the invariance of 
the uniform measure for the ASEP discussed in Sect.~\ref{Measures}. As a consequence, the 
spatial statistics of $h(x,t)$ for long times is that of a Wiener process with 
''diffusion constant'' $D/4 \nu$ \textit{in space}\footnote{An immediate consequence of (\ref{heightdiff}) is that
typical configurations of $h(x,t)$ are non-differentiable. This is the origin of the ill-posedness of the KPZ equation.} , 
\begin{equation}
\label{heightdiff}
\lim_{t \to \infty} \mathbb{E}[(h(x,t) - h(x',t))^2] = \frac{D}{2\nu} \vert x - x' \vert
\equiv A \vert x - x' \vert.
\end{equation}
This relation holds also on the discrete level, provided $\vert x - x' \vert$ is large
compared to the correlation length of the particle system, and it identifies
the ratio $A = D/2 \nu$ as a property of the invariant measure of the latter;
for the continuous time ASEP we read off the relation $A = \bar \rho
(1 - \bar \rho)$ from (\ref{ASEPrough}), and for the 
discrete time TASEP $A$ can be computed from the transition probabilities 
(\ref{DTASEP1},\ref{DTASEP2}) [see (\ref{ADTASEP})]. It can be seen from the relation (\ref{hu})
[or its discrete analogue (\ref{hdiff})] that 
the height difference correlation function  
is a measure of the fluctuations in the
particle number in the interval between $x$ and $x'$. For this reason $A$ has
been referred to as a (nonequilibrium) compressibility \cite{Hager2001}.

We note for later reference that the KPZ-equation (\ref{KPZ}) can be linearized using the Hopf-Cole transformation 
\begin{equation}
\label{HopfCole}
Z(x,t) = \exp \left[ - \frac{\lambda}{2 \nu} h(x,t) \right],
\end{equation}
which was originally applied to the deterministic Burgers equation \cite{Hopf1950,Cole1951} and rediscovered
in the context of (\ref{KPZ}) by Huse, Henley and Fisher \cite{Huse1985}. Indeed, using (\ref{KPZ})
we see that $Z(x,t)$ evolves according to a heat equation with a multiplicative stochastic force,
\begin{equation}
\label{Z(x,t)}
\frac{\partial Z}{\partial t} = \nu \nabla^2 Z +  \frac{\lambda}{2 \nu} \zeta(x,t) Z(x,t).
\end{equation}
The formal solution of (\ref{Z(x,t)}) is a Wiener path integral
describing the weight $Z(x,t)$ of a Brownian path (or ``directed polymer''
\cite{Kardar1986}) subject to the random space-time potential $\frac{\lambda}{2 \nu} \zeta(x,t)$.

\subsection{The universality hypothesis}

The considerations in the preceding subsection suggest that the details of the underlying particle system 
enter the large scale fluctuations properties only through the two parameters $\lambda$
and $A$. These parameters define characteristic scales of height, length and time,
which can be used to non-dimensionalize any correlation function of interest. In the
non-dimensional variables the correlation functions are then conjectured
to be \textit{universal}, i.e.
independent of the specific microscopic model. This is the essence of the 
universality hypothesis. 

As an illustration, consider the probability distribution
of the height $h(x,t)$ at a given point $x$, corresponding to the 
time-integrated current through a fixed bond in the exclusion process. 
Because of translational invariance,
this cannot depend on $x$, and we have to find a combination of $\lambda$, $A$
and $t$ that has the dimension of $h$. Denoting the dimension of a quantity 
$X$ by $[X]$, we read off from (\ref{KPZ}) that 
$$
[\lambda] = \frac{[x]^2}{[h] [t]}
$$
and from (\ref{heightdiff}) that 
$$
[A] = \frac{[h]^2}{[x]}.
$$
The unique combination with the dimension $[h]$ is 
$(A^2 \vert \lambda \vert t)^{1/3}$, and hence we expect
that the rescaled height fluctuation
\begin{equation}
\label{heightscaled}
\tilde h = \frac{h}{(A^2 \vert \lambda \vert t)^{1/3}}
\end{equation}
should have a universal distribution. For example, the variance of the height
is predicted to be of the form \cite{Krug1992}
\begin{equation}
\label{variance}
\mathbb{E}[h(x,t)^2] - (\mathbb{E}[h(x,t)])^2 = c_2 (A^2 \vert \lambda \vert t)^{2/3}
\end{equation}
with a universal constant $c_2$ which is independent of the specific model
or of model parameters such as the update probability $\pi$ in the dTASEP
(see Remark \ref{KPZ_remark}). Similarly, the unique combination of $\lambda$, $A$ and $t$
that has the dimension $[x]$ of length is 
\begin{equation}
\label{correlation_length}
\ell(t) = (A \lambda^2)^{1/3} t^{2/3},
\end{equation}
which defines the \textit{correlation length} of fluctuations; note that correlations spread
\textit{superdiffusively}, that is, faster than 
$t^{1/2}$ \cite{vanBeijeren1985}. This is in contrast to
the case $\lambda = 0$, where a straightforward solution of
(\ref{KPZ}) shows that the correlation length grows diffusively, and
height fluctuations are Gaussian and of order $t^{1/4}$ \cite{Krug1997}.  
This behavior has been explicitly demonstrated for interacting particle systems
in which the current is a linear (or constant) function of the
density\footnote{It is also possible to construct
situations where the leading order nonlinearity in the expansion of the current is of cubic or higher order
in the density fluctuations, so that $\lambda = 0$ but the problem remains
nonlinear \cite{Derrida1991}. 
Non-rigorous analysis indicates that such nonlinearities are irrelevant (in the sense that
the diffusive behavior is preserved) when of quartic order or higher, while in the cubic
case fluctuations spread weakly superdiffusively as $t^{1/2} (\ln
t)^{1/4}$ \cite{Devillard1992,Binder1994}.} \cite{Ferrari1998,Krug2000a,Balazs2006}. 

As an illustration of these considerations, and for later reference, we compute the scale factor
$A^2 \vert \lambda \vert$ for the dTASEP 
at density $\bar \rho = 1/2$. Taking two 
derivatives of the current function (\ref{JTASEP}) we find
$$
\lambda_{\mathrm{dTASEP}}(1/2) = - \frac{2 \pi}{\sqrt{1 - \pi}}.
$$
To determine the compressibility $A$ we appeal to the equivalence of the
invariant measure (\ref{Ising}) to the equilibrium state of  
the one-dimensional Ising chain\footnote{The one-dimensional Ising
chain is treated in most textbooks on statistical physics, see
e.g. \cite{Plischke2006}.}.
Ising spins $\sigma_i$ are canonically related to the occupation variables
$\eta_i$ by $\sigma_i = 1 - 2 \eta_i = \pm 1$. The transition probabilities
(\ref{DTASEP1},\ref{DTASEP2}) make up the \textit{transfer matrix} 
of the Ising chain, with the
density $\rho$ playing the role of the magnetic field (which vanishes when
$\rho = 1/2$) and the update probability $\pi$ controlling the nearest
neighbor coupling; since $p_{1/2}(0,1) > p_{1/2}(1,1)$ the coupling
is \textit{antiferromagnetic} for $\pi > 0$. Particle number fluctuations
translate to fluctuations of the magnetization, and hence the 
compressibility is proportional to the magnetic susceptibility of the Ising
chain. This can be computed from the free energy per spin, which is proportional to the 
logarithm of the largest eigenvalue of the transfer matrix, by taking
two derivatives with respect to the magnetic field. The final result is
\begin{equation}
\label{ADTASEP}
A_{\mathrm{dTASEP}}(1/2) = \frac{1}{4} \;
\frac{p_{1/2}(1,1)}{p_{1/2}(0,1)} = \frac{1}{4}
\sqrt{1 - \pi},
\end{equation}
and we conclude that
\begin{equation}
\label{scaleDTASEP}
(A^2 \vert \lambda \vert)_{\mathrm{dTASEP}} = \frac{1}{8} \pi \sqrt{1 - \pi}.
\end{equation} 

The early work on KPZ-type processes was mostly concerned with establishing the 
universality of the $t^{2/3}$-scaling of the variance (\ref{variance}) which,
once the role of $A$ and $\lambda$ has been recognized, is essentially a 
consequence of dimensional analysis \cite{Krug1997}. Numerical evidence
of universality in a more refined sense, which encompasses universal
\textit{amplitudes} like $c_2$ in (\ref{variance}), 
was presented in \cite{Krug1992}, where it was also pointed
out that different \textit{universality classes} characterized by the same 
$t^{2/3}$ scaling but different amplitudes may arise
from different initial and boundary conditions. Specifically, three cases
were identified:

\begin{itemize}

\item[I.] \textit{Growth from a flat surface without fluctuations}. In the language
of exclusion processes, this corresponds to an \textit{ordered} 
initial condition of constant density; for example, the case $\bar \rho = 1/2$
is realized by occupying all odd or all even sites of the lattice. 

\item[II.] \textit{Growth from a flat surface with stationary roughness}, in the sense
of (\ref{heightdiff}). This corresponds to starting the exclusion process
in a configuration generated from the invariant measure, e.g. a Bernoulli
initial condition of density $\rho$ for the continuous time ASEP. In this
case the universal fluctuations of interest are visible only if the
density is chosen such that the kinematic wave speed
$c(\rho) = 0$; otherwise they will be masked by
the fluctuations in the initial condition which drift across the observation
point. The drift can be eliminated by moving the observation point
at the kinematic wave speed $c(\rho)$. 
 
\item[III.] \textit{Growth of a cluster from a seed.} For the exclusion process
this corresponds to a step initial condition of the form (\ref{step}) with
$\rho_L > \rho_R$. When $\rho_L > 1/2 > \rho_R$ the relation (\ref{phic}) ensures
that the density at the origin $x = 0$ remains at $\phi = c^{-1}(0) = 1/2$
at all times. As in case II., current fluctuations at other values of $\rho$ can be studied
by moving the observation point along a general characteristic $x/t = \xi$
with $\phi(\xi) = \rho$. 

\end{itemize}

Early attempts to derive refined universal information, such as amplitudes and scaling functions, directly
from the KPZ equation met with limited success \cite{Krug1992,Amar1992,Frey1996}. A full understanding 
of the universality classes of the one-dimensional KPZ equation became available only through the spectacular
developments that were triggered a decade ago by the paper of Baik, Deift and Johansson \cite{Baik99}.
In the next three sections we explain the key steps of this development along the lines of the work of 
Johansson \cite{Jo1}, and return to the broader issue  
of KPZ universality in Sect.~\ref{Universality}.

\section{An exactly solvable model: dTASEP with step initial conditions}
\label{tsec:R}

In this section we begin our discussion of Johansson's result \cite{Jo1}
on the fluctuations of the particle flux for discrete time TASEP  (dTASEP) with step initial data. We formulate the result in Theorem \ref{tsatz:R.1} below and we compare it with the predictions of KPZ theory described in the previous section.

Let us first recall the dTASEP model that has been introduced in Sect.~\ref{jsec3.1} (iii).
We denote the infinitely many particles of the system by integers $j = 0, 1, 2, \ldots$ and their respective
positions at integer times $t = 0, 1, 2, \ldots$ by $x_j(t) \in \mathbb Z$. We assume step initial conditions
$x_j(0)=-j$. Jumps to the right $x_j(t+1)=x_j(t)+1$ are attempted at every time step $t \ge 0$ by all particles
$j \ge 0$ independently with probability $\pi$, but have to be discarded by the exclusion property if at time $t$ the receiving site $x_j(t)+1$ is
occupied by another particle of the system, i.e. if $x_{j-1}(t)=x_j(t)+1$. In this case, particle $j$ remains on its site, $x_j(t+1)=x_j(t)$.

\begin{definition}\label{tdef:R.1}
We denote by $\mathbb P_\pi$ the probability measure on the (total) motion of the particle system that is induced by the
stochastic process described above.
\end{definition}

\begin{figure}
\begin{center}
\resizebox{7.cm}{!} {\includegraphics{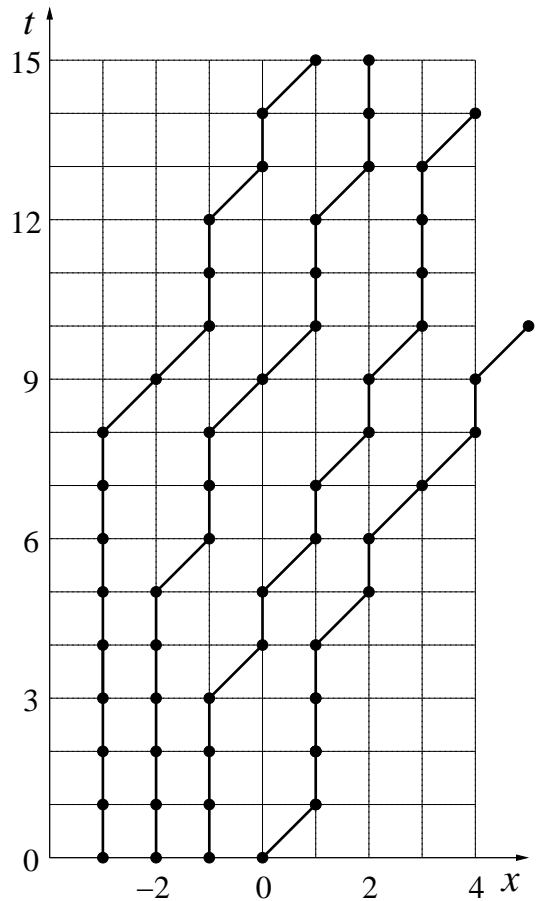}}
\end{center}
\caption{Sample path for dTASEP where only the motion of the four rightmost particles $j=0, 1, 2, 3$ is displayed up to some time $t\le 15$.}
\label{aa}
\end{figure}

Let us first look at an example and compute the probability that the motion depicted
in Figure \ref{aa} occurs. To do this we only need to count for each particle $j = 0, 1, 2, 3$ how many times it had a choice to jump and how often it actually jumped.

\begin{center}
\begin{tabular}{rccc}
      & $\sharp$ choices & $\sharp$ jumps & $\sharp$ stays\\
$j=0$ & $10$       & $5$     & $5$\\
$j=1$ & $11$       & $5$     & $6$\\
$j=2$ & $10$       & $4$     & $6$\\
$j=3$ & $9$       & $4$     & $5$\\
total & $40$      & $18$    & $22$
\end{tabular}
\end{center}

By the assumed stochastic independence of all jumps we have
\[
\mathbb P_\pi \ \textrm{(Figure \ref{aa} occurs)}\ =\pi^{18}(1-\pi)^{22}\;.
\]

Next we turn to the flux which is the quantity that we want to analyze.
\begin{definition}\label{tdef:R.2}
For $r \in \mathbb Z$, $t \in \mathbb N$ we denote the
total flux through the bond between sites $r$ and $r+1$
up to time $t$ by
\[
F_r(t):= \sharp \{ j\in\mathbb N\colon x_j(t)>r\}
-\sharp \{ j\in\mathbb N\colon x_j(0)>r\}\;,
\]
i.e. the total number of particles that have crossed from site $r$ to $r+1$ during the time interval $[0,t]$.
\end{definition}
For example, in the particular situation displayed in Figure \ref{aa} we have

\begin{center}
\begin{tabular}{l|c|c|c|c|}
$t$         & $3$ & $6$ & $9$ & $12$ \\ \hline
$F_{-1}(t)$ & $0$ & $1$ & $2$ & $2$ \\
$F_0(t)$    & $1$ & $2$ & $2$ & $3$ \\
$F_1(t)$    & $0$ & $1$ & $2$ & $2$
\end{tabular}
\end{center}
From now on we will only consider the flux $F_0(t)$ through the bond between sites $0$ and $1$
in order to keep the presentation as simple as possible.

Let us first recall what the discussion on the hydrodynamic limit presented in Sect.~\ref{Hydro}
implies for the current at $x=0$.
We are exactly in the situation of the rarefaction wave (see (ii) of
Sect.~\ref{Hydro}) with $\rho_L = 1$ and $\rho_R =0$ and with $J(\rho)$ given by (\ref{JTASEP}).
Since $c(1/2) = J'(1/2) = 0$ we learn from (\ref{wave}) and (\ref{phic}) that
$\rho(0, t) = \phi(0) = c^{-1}(0) = 1/2$ and again by (\ref{JTASEP}) it follows that the current $j(0, t)$
is given by
\begin{eqnarray}\label{teq:R.3}
j(0, t) = J(\rho(0, t)) = \frac{1}{2} (1 - \sqrt{1 - \pi}) =: J_{\pi}\, .
\end{eqnarray}
We therefore expect that $F_0(t)$ is approximately given by $J_{\pi} t$. Indeed, it is a corollary of
Theorem \ref{tsatz:R.1} below that $F_0(t)/t$ converges with probability 1 to $J_{\pi}$ as $t \to \infty$.

Now we turn to the more detailed predictions of KPZ theory. As it was explained in Sect.~\ref{Growthmodel}, the flux $F_0(t)$ corresponds to the height $h(1/2, t)$. The dimensional analysis of Sect.~\ref{KPZ_conjecture} suggests that the fluctuations of $F_0(t) - J_{\pi}t$
are of order $t^{1/3}$ (see (\ref{heightscaled})). Moreover, taking
(\ref{variance}) and (\ref{scaleDTASEP}) into account, it is predicted for large values of $t$ that the standard deviation of the centered and rescaled variable
$Z_{\pi}(t) := t^{-1/3}(F_0(t) - J_{\pi}t)$ is given by
\begin{eqnarray*}
\sqrt{\mbox{Var} (Z_{\pi}(t))} = C \pi^{1/3} (1-\pi)^{1/6}, \quad t \mbox{ large, }
\end{eqnarray*}
with $C$ independent of $\pi$. The result of Johansson forcefully reaffirms the KPZ conjecture for dTASEP. Even more is true: In the limit $t \to \infty$, the random variables $Z_{\pi}(t)/(\pi^{1/3} (1-\pi)^{1/6})$ do not only have the same second moments for all $0 < \pi < 1$, but they converge to exactly the same probability distribution.

One may think of this result in analogy to the Central Limit Theorem. There one considers
independent, identically distributed random variables $X_i$. The quantities for which we draw the
analogy to the fluxes $F_0(n)$ are the partial sums $S_n = X_1 + \ldots + X_n$. Under some weak assumptions
on the distribution of the $X_i$'s one has with probability 1 that $S_n/n$ converges to the expectation
$\mu := \mathbb E(X_1)$ for $n \to \infty$ (law of large numbers) and that the rescaled random variables
$n^{-1/2}(S_n - n \mu)$ tend to a Gaussian distribution (Central Limit Theorem).

In contrast to the Central Limit Theorem the rescaled variables $Z_{\pi}$ do not converge to a Gaussian distribution. Indeed, and quite surprisingly, the limiting distribution is given by the Tracy-Widom distribution of random matrix theory.

\begin{reminder}\label{trem:C.1}
(Tracy-Widom distribution)

\noindent
{\em
The Gaussian Unitary Ensemble GUE is defined as a sequence $\mathbb P_N$ of Gaussian probability
measures on $N\times N$ Hermitean matrices of the form
\[
d\mathbb P_N(M)=\frac{1}{Z_N}e^{-\mathrm{tr}(M^2)}dM
\]
where $Z_N$ denotes the norming constant and $dM$ abbreviates the product Lebesgue measure on the real and imaginary parts of the entries of $M$ respecting the Hermitean symmetry, i.e. $dM = \Pi_i \,dM_{ii} \,\,
\Pi_{i<j} \,d\mbox{Re}(M_{ij})\, d\mbox{Im}(M_{ij})$. Denote
by $\lambda_1(M)$ the largest eigenvalue of $M$ which is a random object. $\lambda_1(M)$ is expected to be at $\sqrt{2N}$ and the fluctuations are of order $N^{-1/6}$. One can prove that the appropriately rescaled largest eigenvalue converges to some distribution that is now called the Tracy-Widom distribution, i.e. for
$s\in \mathbb R$ we have
\[
\mathbb P_N\left(\frac{\lambda_1(M)-\sqrt{2N}}{(8N)^{-1/6}}\le s\right)\longrightarrow
TW_2(s) \qquad \mbox{as }N\to\infty\, .
%=\int\limits^s_{-\infty} \sigma_2(t)\, dt
\]
The function $TW_2$ can be expressed in terms of the Hastings--McLeod
solution of the Painlev\'e II equation \cite{TTracyWidom94} or by Fredholm determinants of integral operators
with Airy kernel (see Sect.~\ref{tsec:A.2} for more details). Note that the subindex $2$ of $TW_2$ is related
to the fact that GUE is a $\beta$-random matrix ensemble with $\beta=2$. Roughly speaking a $\beta$-ensemble
is an ensemble where the joint distribution of eigenvalues is of the form
\[
d\mathbb P_N(\lambda_1, \ldots, \lambda_N)= \frac{1}{\hat{Z}_N} |\Delta(\lambda)|^{\beta} \prod_{j=1}^N w_N(\lambda_j) d\lambda_j \, ,
\]
and $\Delta$ denotes the Vandermonde determinant (cf. Sect.~\ref{tsec:C.4} below).
See \cite{TMehta} for a general reference on Random Matrix Theory.
The densities of the Tracy-Widom distributions $TW_1$ and $TW_2$ are displayed in Fig.~\ref{3functions}(a) of Sect.~\ref{Universality}.
}
\end{reminder}

For our model, dTASEP with step initial data, the theorem of Johansson \cite{Jo1} implies:

\begin{theorem} \label{tsatz:R.1}
Let $0 < \pi < 1$. Set $J:=\frac{1}{2}(1-\sqrt{1-\pi})$ and $V:= 2^{-4/3} \pi^{1/3}(1-\pi)^{1/6}$. Then, for all $s \in
\mathbb R$ we have
\begin{eqnarray}\label{tN2.5}
\lim_{t \to \infty} \mathbb P_\pi \left(\frac{F_0(t)-Jt}{Vt^{1/3}}\le s\right) =
1-TW_2(-s)
%=\int^\infty_{-s}\sigma_2 (t)\,dt\;.
\end{eqnarray}
\end{theorem}

\begin{remark}
\label{KPZ_remark}
As it was noted above Theorem \ref{tsatz:R.1} affirms and strengthens the KPZ predictions. In particular, the scaling of the flux is precisely that expected from KPZ theory. Comparison with
(\ref{scaleDTASEP}) shows that $V = (A^2 \vert \lambda \vert/2)^{1/3}$ in the notation of Sect.~\ref{KPZ_conjecture}.

\end{remark}

The results of Johansson in \cite{Jo1} are more general than stated above. Mean and fluctuations of the particle flux are described not only at the origin and continuous (in time) TASEP is also considered
by letting $\pi$ tend to $0$ and by rescaling time in an appropriate manner.

We would like to emphasize that Johansson's proof of Theorem \ref{tsatz:R.1} does not make any use of the
considerations regarding the hydrodynamic limit and the KPZ conjecture as presented above. Instead,
the problem treated in Theorem \ref{tsatz:R.1} should be viewed as a very special one within the class of models considered in Sects.~\ref{Exclusion}
and \ref{KPZ_conjecture}. This problem has the attractive feature that it is exactly solvable by a series of
beautiful and non-obvious observations which will be described in the following two sections.

\section{Proof of Theorem \ref{tsatz:R.1} -- part I: Combinatorics}
\label{tsec:C}

We begin our discussion of the proof of Theorem \ref{tsatz:R.1} by relating the flux
$F_0(t)$ to another random variable. For $j,k\in \mathbb N$ denote
\[
T(j,k):= \min \{ t\in\mathbb N\colon x_j(t)=k+1-j\}\;,
\]
that is the time by which particle $j$, that starts at site $x_j(0)=-j$, has just completed its
$(k+1)$-st jump. Observe, that at time $T_k:=T(k,k)$ we have
\[
x_0(T_k)>x_1(T_k)>\ldots >x_k(T_k)=1>0\ge x_{k+1} (T_k)>\ldots\;.
\]
Thus, at time $T_k$ exactly the first $k+1$ particles $0,1,\ldots ,k$ have already jumped from
site $0$ to site $1$ and $F_0(T_k)=k+1$. Moreover, for times $t<T_k$ we have
$F_0(t)\le k$. This implies the relation
\begin{equation}\label{teq:R.9}
\mathbb P_\pi (F_0(t)\le k)=\mathbb P_\pi (T_k>t)=1-\mathbb P_\pi (T(k,k)\le t)\;.
\end{equation}

In the present section we outline how the explicit formula (\ref{teq:C.28}) of Lemma \ref{tlemma:C.1} below for the probability
distribution of $T(k,k)$ can be derived. By a series of bijections we map our combinatorial model
via waiting times and random words to Semi Standard
Young Tableaux, a classical object of combinatorics and representation theory where explicit formulas
for counting are available. The asymptotic analysis of formula
(\ref{teq:C.28}) for $\mathbb P_\pi (T(k,k)\le t)$
is discussed in Sec.~\ref{tsec:A}.

\subsection{From discrete TASEP to waiting times}
\label{tsec:C.1}

We introduce an equivalent description of the dynamics of the particle system by a table of waiting times.
For $j,l\in\mathbb N$ we denote
\begin{eqnarray*}
w_{j,l} &:=& \mbox{ number of times particle $j$ decides to stay on site $l-j$ } \\
&& \mbox{ after it becomes possible to jump to site $l-j+1$. }
\end{eqnarray*}
For example, in the case of Figure \ref{aa} we can determine the following entries of the matrix $(w_{j,l})$ of waiting times.
\begin{equation}\label{teq:C.0}
\left(
\begin{array}{cccccc}
     0&3&1&0&1&\ldots \\
     2&0&0&1&3&\ldots \\
     1&2&0&2&?&\ldots \\
     2&0&2&1&?&\ldots \\
     \vdots&\vdots&\vdots&\vdots&\vdots&\ddots
\end{array}
\right)
\end{equation}
The key observation for computing $T(k,k)$ from the table of waiting times is the
following recursion for $T(j,k)$.
\begin{eqnarray}\label{teq:C.1}
\hspace{-30pt}
    T(j,k)=1+w_{j,k}+\left\{\begin{array}{lcl}
         0&,&\textrm{if}\ j=k=0\\
         T(j,k-1)&,&\textrm{if}\ j=0, k>0\\
         T(j-1,k)&,&\textrm{if}\ j>0, k=0\\
         \max (T(j-1,k), T(j,k-1))&,&\textrm{if}\ j,k>0
     \end{array}\right.
\end{eqnarray}
Indeed, to compute the time it takes the $j$-th particle to complete its $(k+1)$-st jump one needs to add
$1+w_{j,k}$ to the time when this jump became possible. For this jump to become possible, particle $j$ has to be on site $k-j$ (happens at
$T(j, k-1)$) and particle $j-1$ must have emptied neighboring site $k-j+1$ (happens at time $T(j-1,k)$). It is obvious from (\ref{teq:C.1}) that in order to compute $T(k, k)$ one only needs to know the $(k+1)\times (k+1)$ topleft section of the table of waiting times
$(w_{j,l})_{0\le j,l \le k}$.

Relation (\ref{teq:C.1}) allows to prove the following formula for $T(j,k)$ by induction on $(j+k)$:
\begin{eqnarray}\label{teq:C.2}
     T(j,k)=j+k+1+\max\limits_{\mathcal P\in \Pi_{j,k}}\left(\sum\limits_{s
      \scriptsize{\mbox{ on }} \mathcal P}w_s\right)\;.
\end{eqnarray}
Here $\Pi_{j,k}$ denotes the set of paths $\mathcal P$ in the table of waiting times that connect the
$(0,0)$-entry with the $(j,k)$-entry and satisfy the additional condition that only steps to the right-neighbor and to the
neighbor downstairs are permitted. More formally, we may write
\begin{eqnarray*}
\Pi_{j,k}=\left\{ (s_0,\ldots ,s_{j+k})\in(\mathbb N\times\mathbb N)^{j+k+1}\colon
s_0=(0,0), s_{j+k}=(j,k)\ \textrm{and}\right.\\
\left. s_i-s_{i-1} \in \{(1,0),(0,1)\}\ \textrm{for all}\ 1\le i\le j+k\right\}\;.
\end{eqnarray*}
For $\mathcal P =(s_0,\ldots ,s_{j+k})\in \Pi_{j,k}$ we understand
\[
\sum\limits_{s \scriptsize{\mbox{ on }} \mathcal P}w_s:= \sum\limits^{j+k}_{i=0}w_{s_i}
\]
We illustrate formula (\ref{teq:C.2}) with our running example. The corresponding table of waiting times
displayed in (\ref{teq:C.0}) has nine paths in $\Pi_{3,3}$
that maximize the sum of waiting times. Two of them are
\begin{eqnarray}\label{teq:C.N1}
     \mathcal P_1 &\colon & (0,0)\to (0,1)\to (0,2)\to (0,3)\to (1,3)\to (2,3)\to (3,3)\\
     \mathcal P_2&\colon & (0,0)\to (1,0)\to (2,0)\to (2,1)\to (2,2)\to (3,2)\to (3,3)
\end{eqnarray}
and we have
\[
\sum\limits_{s \scriptsize{\mbox{ on }} \mathcal P_1}w_s=\sum\limits_{s \scriptsize{\mbox{ on }} \mathcal P_2}w_s=8\;.
\]
Formula (\ref{teq:C.2}) then yields for the time when the fourth particle $j=3$ has just completed its fourth jump $T(3,3)=3+3+1+8=15$ which is easily verified from Figure \ref{aa}.

\begin{remark}
The probabilistic model we have arrived at, i.e. to search for right- and downward
paths that maximize the total waiting time, is also known as the last passage
percolation problem and that is precisely the model studied in the paper
\cite{Jo1} of Johansson. Interpreting $w_{j,l}$ as potential energies this can also be considered as the problem
of zero-temperature directed polymers in a random medium \cite{Kardar1987,Krug1992,Krug1994,HalpinHealy1995,Krug1998}.
\end{remark}

We are now ready to compute $P_{\pi} (T(k,k) \le t)$ in terms of the table of waiting times. Recall from (\ref{teq:C.2}) that $T(k, k)$ is completely determined by the topleft $(k+1) \times (k+1)$ corner of the table. Moreover, and again by (\ref{teq:C.2}), a table of waiting times corresponds to a particle dynamics with $T(k,k) \le t$ if and only if the topleft corner belongs to the set $W(k, t)$
which we define to be the set of $(k+1) \times (k+1)$ matrices $(w_{j,l})$ with entries that are non-negative integers
and which have the additional property
\begin{eqnarray}\label{teq:C.4}
\max_{\mathcal P\in \Pi_{k,k}}\left(\sum_{s \scriptsize{\mbox{ on }} \mathcal P}w_s\right) \le t - 2k - 1 \; .
\end{eqnarray}

It is straightforward to determine the probability that the topleft corner of the table of waiting times agrees with any given element $Q$ of $W(k, t)$. Indeed, we only need to count the total number of decisions
to either jump (always equals $(k+1)^2$) or to stay (equals the sum of all entries of $Q$ which we denote by $|Q|_1$) whenever a jump is not prohibited by the exclusion property. In summary
we obtain the following formula
\begin{eqnarray}\label{teq:C.3}
\mathbb P_{\pi} (T(k,k) \le t) = \sum_{Q \in W(k,t)} \pi^{(k+1)^2} (1-\pi)^{|Q|_1}.
\end{eqnarray}

\subsection{From waiting times to random words}
\label{tsec:C.2}

We associate with any $(k+1)\times (k+1)$ matrix $Q = (w_{j,l})$ of waiting times the sequence of pairs
$(j,l)_{0 \le j, l \le k}$, listed in lexicographical order, where the value of $w_{j,l}$ determines how often the index
$(j, l)$ appears in this list. In the case of $Q$ being the top left $4 \times 4$ submatrix in (\ref{teq:C.0}) the corresponding sequence of pairs reads
\begin{eqnarray}\label{teq:C.10}
\begin{array}{ccccccccccccccccc}
0&0&0&0&1&1&1&2&2&2&2&2&3&3&3&3&3\\
1&1&1&2&0&0&3&0&1&1&3&3&0&0&2&2&3
\end{array}
\end{eqnarray}
We may consider this list of pairs as a list of 17 two-letter words from the alphabet $\{0, 1, 2, 3\}$ in lexicographical order. This
explains the term ``random words'' often used in this context. Observe that any right-downward path $\mathcal P\in \Pi_{3,3}$ corresponds to a subsequence in this list of 17 two-letter words where both the first and the second row are weakly increasing.
A little more thought shows that the quantity $\max_{\mathcal P\in \Pi_{3,3}}\left(\sum_{s \scriptsize{\mbox{ on }} \mathcal P}w_s\right)$
is given by the length of the longest subsequence that is
weakly increasing in both rows. The sequence (\ref{teq:C.10}) has nine such subsequences of maximal length 8. The subsequences corresponding to the paths $\mathcal P_1$ and $\mathcal P_2$ of (\ref{teq:C.N1}) read
\begin{eqnarray*}
\begin{array}{cccccccc}
0&0&0&0&1&2&2&3\\
1&1&1&2&3&3&3&3
\end{array}, \qquad \mbox{ and } \qquad
\begin{array}{cccccccc}
1&1&2&2&2&3&3&3\\
0&0&0&1&1&2&2&3
\end{array} .
\end{eqnarray*}
Formula (\ref{teq:C.3})
translates to
\begin{eqnarray}\label{teq:C.12}
\mathbb P_{\pi} (T(k,k) \le t) = \sum_{\phi \in D(k,t)} \pi^{(k+1)^2} (1-\pi)^{\mbox{length of } \phi},
\end{eqnarray}
where $D(k, t)$ is the set of finite sequences $\phi$ of lexicographically ordered two-letter words from the
alphabet $\{0, 1, \ldots, k \}$ and for which the length of the longest subsequence of $\phi$ that increases weakly in both letters
is at most $t-2k-1$. By the Robinson--Schensted--Knuth correspondence we may enumerate the set $D(k, t)$
conveniently in terms of Semi Standard Young Tableaux. This is the content of the next section.

\subsection{From random words to Semi Standard Young Tableaux}
\label{tsec:C.3}

The Robinson--Schensted correspondence provides a bijection between permutations and Standard Young Tableaux that
is well known in combinatorics and in the representation theory of the permutation group. We now describe the extension
of this algorithm to random words which was introduced by Knuth \cite{Knu}. The basic algorithm that needs to be understood first
is the row insertion process. Suppose we have a weakly increasing sequence of integers, e.g. $0 \; 0 \; 1 \; 1 \; 1 \; 3$.
We insert an integer $r$ into this row by the following set of rules. If $r \ge 3$ we simply append $r$ at the end of the row.
In the case $r < 3$ we replace the unique number $s$ in the row that is strictly bigger than $r$ such that after the replacement
the sequence is still weakly increasing. We say that we have inserted $r$ by bumping $s$. For the sequence
$0 \; 0 \; 1 \; 1 \; 1 \; 3$ insertion of $r$ leads to
\begin{center}
\begin{tabular}{|l|l|l|}
$r$         & sequence after insertion of $r$ & bumped number \\ \hline
$0$ & $0 \; 0 \; 0 \; 1 \; 1 \; 3$ &$1$ \\
$1$ & $0 \; 0 \; 1 \; 1 \; 1 \; 1$ & $3$ \\
$2$ & $0 \; 0 \; 1 \; 1 \; 1 \; 2$ & $3$ \\
$3$ & $0 \; 0 \; 1 \; 1 \; 1 \; 3 \; 3$ & no number bumped
\end{tabular}
\end{center}
We now describe the procedure how a lexicographically ordered list of two-letter random words is transformed
into a pair of tableaux. In a first step, one only considers the sequence of the second letters of the random words
in order to build the first tableau. To this end
one inserts (see row insertion process described above) the second letters of the words, one after the other, into the first row of the tableau
that has been created before. In case a number is bumped, the bumped number is inserted into the second row of the tableau.
In case bumping occurs again in the second row, we insert the newly bumped number into the third row of the tableau.
This process is repeated until the bumping ends. In our example (\ref{teq:C.10}) this process leads to the following
sequence of tableaux.
\begin{equation*}
\hspace{-20pt}
\begin{array}{c}
1
\end{array}
\stackrel{1}{\longrightarrow}
\begin{array}{cc}
1&1
\end{array}
\stackrel{1}{\longrightarrow}
\begin{array}{ccc}
1&1&1
\end{array}
\stackrel{2}{\longrightarrow}
\begin{array}{cccc}
1&1&1&2
\end{array}
\stackrel{0}{\longrightarrow}
\begin{array}{cccc}
0&1&1&2\\
1&&&
\end{array}
\stackrel{0}{\longrightarrow}
\end{equation*}
\begin{equation*}
\hspace{-20pt}
\begin{array}{cccc}
0&0&1&2\\
1&1&&
\end{array}
\stackrel{3}{\longrightarrow}
\begin{array}{ccccc}
0&0&1&2&3\\
1&1&&&
\end{array}
\stackrel{0}{\longrightarrow}
\begin{array}{ccccc}
0&0&0&2&3\\
1&1&1&&
\end{array}
\stackrel{1}{\longrightarrow}
\end{equation*}
\begin{equation*}
\hspace{-20pt}
\begin{array}{ccccc}
0&0&0&1&3\\
1&1&1&2&
\end{array}
\stackrel{1}{\longrightarrow}
\begin{array}{ccccc}
0&0&0&1&1\\
1&1&1&2&3
\end{array}
\stackrel{3}{\longrightarrow}
\begin{array}{cccccc}
0&0&0&1&1&3\\
1&1&1&2&3&
\end{array}
\stackrel{3}{\longrightarrow}
\end{equation*}
\begin{equation*}
\hspace{-20pt}
\begin{array}{ccccccc}
0&0&0&1&1&3&3\\
1&1&1&2&3&&
\end{array}
\stackrel{0}{\longrightarrow}
\begin{array}{ccccccc}
0&0&0&0&1&3&3\\
1&1&1&1&3&&\\
2&&&&&&
\end{array}
\stackrel{0}{\longrightarrow}
\end{equation*}
\begin{equation*}
\hspace{-20pt}
\begin{array}{ccccccc}
0&0&0&0&0&3&3\\
1&1&1&1&1&&\\
2&3&&&&&
\end{array}
\stackrel{2}{\longrightarrow}
\begin{array}{ccccccc}
0&0&0&0&0&2&3\\
1&1&1&1&1&3&\\
2&3&&&&&
\end{array}
\stackrel{2}{\longrightarrow}
\end{equation*}
\begin{equation*}
\hspace{-20pt}
\begin{array}{ccccccc}
0&0&0&0&0&2&2\\
1&1&1&1&1&3&3\\
2&3&&&&&
\end{array}
\stackrel{3}{\longrightarrow}
\begin{array}{cccccccc}
0&0&0&0&0&2&2&3\\
1&1&1&1&1&3&3&\\
2&3&&&&&&
\end{array}
\end{equation*}
Using this procedure we have obtained a Semi Standard Young Tableau with 17 entries.
\begin{definition}\label{tdef:C.1}
By a Semi Standard Young Tableau (SSYT) we understand a tableau $\cal T$ of a finite number of integers
that are weakly increasing in each row and strictly increasing in each column. The shape $\lambda$ = sh$(\cal T)$
of $\cal T$ is denoted by the sequence of row lengths $(\lambda_0, \lambda_1, \ldots)$ that is required to be
a weakly decreasing sequence of non-negative integers. Furthermore, we set $|\lambda|:= \sum_i \lambda_i$ to be
the total number of cells in the tableau.
\end{definition}
In order to see that the procedure described above always leads to a SSYT one only needs to convince oneself that adding one number to
a SSYT will result in a SSYT with one more cell. The key observation here is that in the bumping procedure a number can only move downwards or left-downwards.

In our running example the final SSYT $\cal T^*$ has shape
$(8,7,2,0,0,\ldots)$ and it is no coincidence but a theorem that the length of the first row $\lambda_0$ equals the length of the longest weakly increasing subsequence that we have seen to be 8. This fact can be proven in general by showing inductively that at every step of the procedure the length of the longest weakly increasing subsequence (in both letters) up to some word in the list is given by the position at which the second letter of the word is inserted in the first row of the tableau. In our example (\ref{teq:C.10}) the length of the longest weakly increasing subsequence up to the 15-th word $(3,2)$ is six and correspondingly $2$ is inserted at the sixth position of the first row.

Note that the sequence (\ref{teq:C.10})
is not the only one that leads to the final tableau $\cal T^*$. For example, the sequence
\[
\begin{array}{ccccccccccccccccc}
0&0&0&0&1&1&1&2&2&2&2&2&3&3&3&3&3\\
1&1&1&2&3&3&3&0&0&1&1&3&0&0&0&2&2
\end{array}
\]
leads to the same $\cal T^*$. However, and this is the central message of the Robinson-Schensted-Knuth correspondence, one may encode the sequence of random words (\ref{teq:C.10}) in a unique way
if one records in addition how the tableau grows and if one remembers the first letters of the random words that have so far been neglegted.
This information is all encoded in the second tableau. We now demonstrate how to build this second tableau
with our running example. As a first step we record for the above described procedure which cell has been
added to the SSYT at which step.
\begin{equation}\label{teq:C.16}
\begin{array}{cccccccc}
1&2&3&4&7&11&12&17\\
5&6&8&9&10&15&16&\\
13&14&&&&&&
\end{array}
\end{equation}
Since we also need to remember the first letters of our 17 random words it is natural to replace
the entries in (\ref{teq:C.16}) in the following way. We note that the first 4 words in (\ref{teq:C.10}) have first letter 0 and we therefore
replace $1$, $2$, $3$, $4$ each by $0$. The next three words have first letter 1 and we replace $5$, $6$, $7$ each by $1$. Then there are five words starting with letter $2$, leading us to replace $8$, $9$, $10$, $11$, $12$ each by $2$.
The remaining five entries $13$, $14$, $15$, $16$, $17$ are each replaced by $3$. This leads to the tableau
\begin{equation}\label{teq:C.18}
\begin{array}{cccccccc}
0&0&0&0&1&2&2&3\\
1&1&2&2&2&3&3&\\
3&3&&&&&&
\end{array}\, .
\end{equation}
We have arrived at a
SSYT $\cal U^*$ that clearly has
the same shape as $\cal T^*$. In order to prove that the second tableau always yields a SSYT there is only one non-obvious property to verify, which is the strict increase in each column. To see this recall that for words $(a_1, a_2)$,
$(b_1, b_2)$ the number $b_2$ can only bump $a_2$ if $a_2 > b_2$ which implies by the lexicographical ordering of the list that $b_1 > a_1$. Therefore, in the second tableau, $b_1$ will be located below or left-below $a_1$ and by the weak increase within each row this suffices.

In summary we have mapped the sequence (\ref{teq:C.10}) of random words to the pair of SSYT's
$(\cal T^*, \cal U^*)$
\begin{equation*}
\begin{array}{cccccccc}
0&0&0&0&0&2&2&3\\
1&1&1&1&1&3&3&\\
2&3&&&&&&
\end{array}
\;, \qquad \qquad
\begin{array}{cccccccc}
0&0&0&0&1&2&2&3\\
1&1&2&2&2&3&3&\\
3&3&&&&&&
\end{array}
\end{equation*}
of equal shape $\lambda$. It is an
instructive exercise to reconstruct sequence (\ref{teq:C.10}) from the pair of SSYT's. In fact,
the proof that the above described procedure maps lexicographically ordered two letter words bijectively to pairs of SSYT's of equal shape
can be given by an explicit description of the inverse map. This bijection has two more features that are of interest to us.
Firstly, $|\lambda|$ equals the number of words in our list (= 17 in our running example). Secondly,
the length of a longest weakly increasing subsequence is exactly given by the length of the first row $\lambda_0$
(= 8 in our example) as we have already observed above (see \cite{Knu} for details).

This implies that the set $D(k, t)$ (cf. (\ref{teq:C.12})) is bijectively mapped onto the set of all pairs
$(\cal T, \cal U)$ of SSYT's of equal shape $\lambda$ satisfying
\[
t-2k-1 \ge \lambda_0 \ge \lambda_1 \ge \ldots
\]
with entries from $\{ 0, 1, \ldots, k \}$. Note that we have $\lambda_{k+1}=0$ because entries in each column are strictly increasing. We therefore arrive at
\begin{equation}\label{teq:C.20}
\mathbb P_{\pi} (T(k,k) \le t) = \sum_{t-2k-1 \ge \lambda_0 \ge \ldots \ge \lambda_{k} \ge 0} \pi^{(k+1)^2} (1-\pi)^{| \lambda |} L(\lambda, k)^2,
\end{equation}
where $L(\lambda, k)$ denotes the number of SSYT's of shape $\lambda = (\lambda_0, \ldots, \lambda_{k}, 0, \ldots)$
and with entries from $\{0, 1, \ldots, k \}$. We have now derived a representation for $\mathbb P_{\pi} (T(k,k) \le t)$
involving the combinatorial quantity $L(\lambda, k)$ that can be computed explicitly.

\subsection{Schur polynomials and an explicit formula for the distribution of $T(k, k)$}
\label{tsec:C.4}

There is a beautiful argument using Schur polynomials $s_{\lambda}$
that provides a formula for $L(\lambda, k)$. This argument is
explained in the appendix of \cite{Sas1} and we do not repeat it
here (see in addition \cite[Cor. 4.6.2]{TSagan01} for a derivation
of the representation \cite[(A.5)]{Sas1} of $s_{\lambda}$ by
determinants). The result is
\begin{equation}\label{teq:C.22}
L(\lambda, k) = \prod_{0\le i < j \le k} \frac{\lambda_i - \lambda_j + j - i}{j-i}.
\end{equation}
Introducing the new variables $y_i := \lambda_i -i +k$ and denoting the Vandermonde determinant by $\Delta(y) = \prod_{0\le i < j \le k}
(y_j-y_i)$  we obtain
\begin{equation}\label{teq:C.24}
\hspace{-30pt}
\mathbb P_{\pi} (T(k,k) \le t) = C_{\pi, k}  \sum_{t-k-1 \ge y_0 > \ldots > y_{k} \ge 0}
\Delta(y)^2 \prod_{i=0}^k (1-\pi)^{y_i}, \quad \mbox{ where }
\end{equation}
\begin{equation} \label{teq:C.26}
C_{\pi, k} := \pi^{(k+1)^2} (1-\pi)^{-k(k+1)/2}
\prod_{0\le i < j \le k} \frac{1}{(j-i)^2}
\end{equation}
Observe that each term in the sum is a symmetric function in $y$ that vanishes if two components agree.
This leads to the final formula in this section for the probability distribution of $T(k, k)$ which is closely related to the distribution of the flux $F_0(t)$ via (\ref{teq:R.9}).
\begin{lemma}\label{tlemma:C.1}
\begin{eqnarray}\label{teq:C.28}
\mathbb P_{\pi} (T(k,k) \le t) &=& \frac{C_{\pi, k}}{(k+1)!}
\sum_{\scriptsize{\begin{array}{c} y \in \mathbb Z^{k+1} \\ 0 \le y_i \le t-k-1\end{array}}}
\Delta(y)^2 \prod_{i=0}^k (1 - \pi)^{y_i}\; .
\end{eqnarray}
\end{lemma}
This formula should be compared with the formula for the distribution of the largest eigenvalue of
the Gaussian Unitary Ensemble (cf. Reminder \ref{trem:C.1})
\begin{eqnarray}\label{teq:N.20}
\mathbb P_N (\lambda_1(M) \le \Lambda) = \frac{1}{\hat{Z}_N} \int_{(-\infty, \Lambda]^N} \Delta(y)^2
\prod_{j=1}^N e^{-y_j^2} \ dy
\end{eqnarray}
with some appropriate norming constant $\hat{Z}_N$. Observe that
this formula has exactly the same structure as (\ref{teq:C.28}). The
role played by the measure $e^{-x^2}dx$ for GUE is taken by the
discrete measure $\sum_{j=0}^{\infty} (1-\pi)^j \delta_j$ supported
on $\mathbb N$ for dTASEP. Here $\delta_j$ denotes the $\delta$-distribution concentrated at $j$. In the next section we discuss how the
method of orthogonal polynomials can be used to analyze the
asymptotics of such types of high-dimensional integrals.

\section{Proof of Theorem \ref{tsatz:R.1} -- part II: Asymptotic analysis}
\label{tsec:A}

In the previous section we have derived formulae (\ref{teq:C.28}) and (\ref{teq:R.9}) for the probability distribution of the flux $F_0(t)$.
We now need to analyze this formula asymptotically in a regime where $t$ and $k$ both become large and $k \approx J t + V t^{1/3} s$ with $s$ being an arbitrary but fixed real number and $J$, $V$ being defined as in Theorem \ref{tsatz:R.1}. The key to the analysis is the observation that the right hand side of (\ref{teq:C.28}) is structurally
the same as the standard formula for the probability distribution of the largest eigenvalue of GUE (\ref{teq:N.20}) and
the method of orthogonal polynomials (see Sect.~\ref{tsec:A.1}) can be applied. The role played by Hermite polynomials for GUE will
be taken by Meixner polynomials in our model. In both cases it is the behavior of the
orthogonal polynomials of large degree in a vicinity of their respective largest zero that matters in the
asymptotic analysis.
After appropriate rescaling this behavior can be described in terms of Airy functions for
both Hermite and Meixner polynomials (see Sect.~\ref{tsec:A.2}). On a technical level this explains the occurence of the Tracy-Widom distribution $TW_2$ for GUE as well as for dTASEP with step initial conditions. We include in Sect.~\ref{tsec:A.3} a brief discussion of the universal behavior of orthogonal polynomials.

\subsection{The method of orthogonal polynomials following an approach of Tracy and Widom}
\label{tsec:A.1}

Almost 50 years ago, Gaudin and Mehta have introduced orthogonal polynomials to random matrix theory in order to study local eigenvalue statistics. A very transparent version of the now so called \emph{method of orthogonal polynomials} is due to
Tracy and Widom \cite{TW2} and we will briefly outline their approach for our situation. See also the recent book of Deift and Gioev \cite[Chapt. 4]{TDeiftGioev09} for a self contained presentation of the method of orthogonal polynomials with some remarks on the history of the method. For us the method allows to express the right hand side of (\ref{teq:C.28}) in terms of a Fredholm determinant of an operator where the (discrete) integral kernel is given in terms of Meixner orthogonal polynomials.

We begin our discussion by introducing the discrete weight
\[
w_{\pi}(x) := \left\{
\begin{array}{ll}
0&, \mbox{ if } x < 0 \\ (1-\pi)^x &, \mbox{ if } x \ge 0
\end{array}
\right. , \quad x \in \mathbb Z\; .
\]
For a moment we let $(q_l)_{l\ge 0}$ be any sequence of polynomials with $q_l$ being of degree $l$ with (non-zero) leading coefficient
$\gamma_l$. Setting $\varphi_l (x) := q_l (x) \sqrt{w_{\pi}(x)}$ and using the definition of the Vandermonde determinant together with some basic properties of determinants we have for $y \in \mathbb N^{k+1}$ that
\[
\left[ \det (\varphi_l(y_i))_{0 \le i, l \le k} \right]^2 =
(\gamma_0 \ldots \gamma_k)^2 \Delta(y)^2 \prod_{i=0}^k (1 - \pi)^{y_i} \; .
\]
Furthermore, we set $I_s := [s, \infty)$ and denote by ${\bf 1}_{I_s}$ its characteristic function that
takes the value 1 on $I_s$ and 0 on $\mathbb R \setminus I_s$. We may then rewrite (\ref{teq:C.28}) in the form
\begin{eqnarray*}
\hspace{-30pt}
\mathbb P_{\pi} (T(k,k) \le t) = \frac{C_{\pi, k}}{(\gamma_0 \ldots \gamma_k)^2(k+1)!}
\sum_{y \in \mathbb Z^{k+1}} \left[ \det (\varphi_l(y_i)) \right]^2 \prod_{i=0}^k
\left( 1 - {\bf 1}_{I_{t-k}}(y_i)
\right).
\end{eqnarray*}
An identity due to Andr\'eief (see e.g. \cite[(3.3)]{TDeiftGioev09}) adapted to our context reads
\begin{eqnarray*}
\hspace{-40pt}
\sum_{y \in \mathbb Z^{k+1}} \left[ \det (\varphi_j(y_i)) \right]
\left[ \det (\varphi_l(y_i)) \right] \prod_{i=0}^k
f(y_i) = (k+1)! \det \left( \sum_{x \in \mathbb Z} \varphi_j(x) \varphi_l(x) f(x)
\right).
\end{eqnarray*}
One may prove this formula using the Leibniz sum for determinants. This allows to write the distribution of $T(k, k)$ as an determinant
\[
\mathbb P_{\pi} (T(k,k) \le t) =
\frac{C_{\pi, k}}{(\gamma_0 \ldots \gamma_k)^2} \det S \; ,
\]
where $S$ denotes the $(k+1) \times (k+1)$ matrix with entries
\[
S_{j, l} = \sum_{x \in \mathbb Z} \varphi_j(x) \varphi_l(x)
\left( 1 - {\bf 1}_{I_{t-k}}(x)
\right) \; ,
\qquad 0 \le j, l \le k \; .
\]
So far the choice of the polynomials $q_l$ of degree $l$ was arbitrary. Now we choose $(q_l)_l$ to be the sequence of
normalized orthogonal polynomials with respect to the discrete measure $\sum_{x \in \mathbb Z} w_{\pi}(x) \delta_x$
which belongs to the class of Meixner polynomials. We have
\[
\sum_{x \in \mathbb Z} \varphi_j(x) \varphi_l(x) = \sum_{x \in \mathbb Z} q_j(x) q_l(x) w(x) = \delta_{j, l}
\]
for $j, l \in \mathbb N$. Hence $S = I - R(t-k)$ with
\[
R(s)_{j, l} = \sum_{x \in \mathbb Z} \varphi_j(x) \varphi_l(x) {\bf 1}_{I_s}(x) =
\sum_{x \ge s} \varphi_j(x) \varphi_l(x) \; .
\]
In summary, we have so far derived
\begin{equation}\label{teq:A.10}
\mathbb P_{\pi} (T(k,k) \le t) = \frac{C_{\pi, k}}{(\gamma_0 \ldots \gamma_k)^2} \det (I - R(t-k)) .
\end{equation}
The prefactor
$C_{\pi, k}(\gamma_0 \ldots \gamma_k)^{-2}$ can be seen to equal $1$ by considering equation (\ref{teq:A.10}) for fixed $k$ in the limit $t \to \infty$. Thus
\begin{equation}\label{teq:AN.10}
\mathbb P_{\pi} (T(k,k) \le t) = \det (I - R(t-k)).
\end{equation}
The final idea in the argument of Tracy-Widom is to write $R(s)$ -- considered as a linear map
$\mathbb R^{k+1} \to \mathbb R^{k+1}$ -- as a product $R(s)= A(s)B(s)$, with
\begin{eqnarray*}
B(s): \mathbb R^{k+1} \to \ell_2(\mathbb Z \cap I_s)&,\qquad&(u_j)_{0 \le j \le k} \mapsto \sum_{j=0}^k u_j \varphi_j|_{I_s} \\
A(s): \ell_2(\mathbb Z \cap I_s) \to \mathbb R^{k+1}&,\qquad&f \mapsto
\left( \sum_{x \ge s} f(x) \varphi_{l}(x)  \right)_{0 \le l \le k}
\end{eqnarray*}
Applying the formula $\det(I-AB)=\det(I-BA)$ that holds in great generality (see e.g. \cite[(3.1)]{TDeiftGioev09}) we have derived the
following Fredholm determinant formula for $\mathbb P_{\pi} (T(k,k) \le t)$.
\begin{lemma}\label{tlem:A.1}
$\mathbb P_{\pi} (T(k,k) \le t) = \det(I-\Sigma_k(t-k))$, where
\[
\Sigma_k(s) \colon \ell_2(\mathbb Z \cap I_s) \to \ell_2(\mathbb Z \cap I_s)\;, \quad f \mapsto
\left( \sum_{y \ge s} \sigma_k(x, y) f(y)  \right)_{x \ge s}
\]
and $\sigma_k$ denotes the reproducing kernel $\sigma_k(x, y) := \sum_{j=0}^k \varphi_j(x) \varphi_j (y)$ with respect
to the Meixner polynomials.
\end{lemma}
It may seem somewhat strange to convert (\ref{teq:A.10}) that involves a determinant of some finite size matrix $I-R$
into a formula that involves the computation of a Fredholm determinant of an operator acting on the infinite dimensional space
$\ell_2(\mathbb Z \cap I_s)$.
However, one has to keep in mind that we are interested in an asymptotic result with $k \to \infty$.
Hence the size of $I-R$ goes to infinity and it is not at all clear how to perform the asymptotic analysis of the determinants.
In contrast, the operator
$I - \Sigma_k$ acts on the same space $\ell_2(\mathbb Z \cap I_s)$ for all $k$ and it is only the reproducing kernels $\sigma_k$ that dependend on $k$. As will be discussed below the kernels $\sigma_k$ are amenable to asymptotic analysis. In fact, due to the
Christoffel-Darboux formula for orthogonal polynomials we may express $\sigma_k$ just in terms of $\varphi_{k}$ and
$\varphi_{k+1}$. For large values of $k$ the behavior of these functions is rather well understood.
For example, if $x$ is somewhat larger than the largest zero of $\varphi_k$, then $|\varphi_k(x)|$ is very close to zero.
This implies that for values of $t-k$ that are somewhat larger than the
largest zeros of $\varphi_k$ and $\varphi_{k+1}$ the operator $\Sigma_k(t-k)$ is negligible and
thus $\mathbb P_{\pi} (T(k,k) \le t)$ is very close to 1.
If one reduces the value of $t-k$ to lie in a vicinity of the largest zero of $\varphi_k$ (which is also close to the
largest zero of $\varphi_{k+1}$) then the functions $\varphi_k$ and $\varphi_{k+1}$, appropriately rescaled, are described to leading order by Airy functions. In Sect.~\ref{tsec:A.2} we will use the just mentioned properties of Meixner
polynomials to complete the proof of Theorem \ref{tsatz:R.1}.

As it was noted in the last paragraph of Sect.~\ref{tsec:C.4}, the formula for the distribution of the
largest eigenvalue of GUE (\ref{teq:N.20}) is structurally the same as formula (\ref{teq:C.28}) for the distribution of $T(k, k)$
and the arguments described in the present section can
be applied in an analogous way. The only difference is that we need to use Hermite polynomials instead of
Meixner polynomials and that the summation operator $\Sigma_k$ is to be replaced by an integral operator with a
kernel that is given by the reproducing kernel for Hermite polynomials up to degree $N-1$ ($N$ as in $\mathbb P_N$,
cf. Reminder \ref{trem:C.1}). As in the Meixner case, the leading order behavior of Hermite polynomials near their largest zero is described by Airy functions. On a technical level this is the reason why the fluctuation of
the flux in dTASEP with step initial conditions follows asymptotically the same distribution as the fluctuation of the largest eigenvalue of
GUE. It is no coincidence that Meixner polynomials and Hermite polynomials of large degree look locally the same when rescaled appropriately. In fact, large classes of orthogonal polynomials display the same local behavior. We will comment on this universality property of orthogonal polynomials in Sect.~\ref{tsec:A.3}.

\subsection{Completing the proof of Theorem \ref{tsatz:R.1}}
\label{tsec:A.2}

We have argued above that in order to complete the proof of Theorem \ref{tsatz:R.1} we need to determine the asymptotic behavior of the reproducing kernel $\sigma_k$ (see Lemma \ref{tlem:A.1}) in a vicinity of the largest zero of $\varphi_k$. This requires some detailed analysis that we are not going to present here and we refer the reader to \cite[Sect. 5]{Jo1}. See also Sect.~\ref{tsec:A.3} for a few general remarks on the asymptotic analysis of orthogonal polynomials.
We start with some notation. The Airy function can be defined for $x \in \mathbb R$ by
\begin{eqnarray}\label{tAi.5}
\textrm{Ai}(x) := \frac{1}{2 \pi} \int_{\mathbb R} \exp \big( i[x(t+is)+(t+is)^3/3]\big) dt
\end{eqnarray}
with an arbitrary choice of $s>0$. The Airy kernel is given by
\begin{eqnarray*}
A(x, y) := \frac{\textrm{Ai}(x) \textrm{Ai}'(y) - \textrm{Ai}'(x) \textrm{Ai}(y)}{x-y} \; ,\quad x, y \in \mathbb R\; ,
\end{eqnarray*}
with the obvious interpretation on the diagonal $x=y$. The
Tracy-Widom distribution for $\beta=2$ can then be expressed as a Fredholm determinant
\begin{eqnarray*}
\textrm{TW}_2(s) := \det (I - A)|_{L^2[s, \infty)} \; , \quad s \in \mathbb R ,
\end{eqnarray*}
where $A$ denotes the integral operator associated with the Airy kernel. Note that one may derive a differential equation for $TW_2$  leading to another representation \cite{TTracyWidom94}
\[
\textrm{TW}_2 (s) = \exp \left( - \int_s^{\infty} (x-s) u(x)^2 dx  \right) \; ,
\]
where $u$ denotes the Hastings-McLeod solution of the Painlev\'e II equation $u''= 2u^3 +xu$ that is singled out from all solutions of this ordinary differential equation by the asymptotic condition $u(x) \sim -$Ai$(x)$ for $x \to \infty$.
Observe that the Airy function solves the linearized Painlev\'e II equation $u''=x u$ with asymptotics Ai$(x) \sim
\frac{\exp \left(-(2/3)x^{3/2}\right)}{2 \sqrt{\pi} x^{1/4}}$ as $x \to \infty$.
In an interesting paper \cite{TBornemann09} Bornemann demonstrates that it is advantageous for numerical evaluations of $TW_2$ to start from the Fredholm determinant formula rather than using the Hastings-McLeod function.

The result in \cite[Lemma 3.2]{Jo1} on the reproducing kernel $\sigma_k$ for Meixner polynomials reads
\begin{equation}\label{teq:A.20}
c k^{1/3} \sigma_k( b k + c k^{1/3} \xi, b k + c k^{1/3} \eta) \to A(\xi, \eta) \quad \mbox{ for } k \to \infty,
\end{equation}
where $b= \pi^{-1}(1+\sqrt{1-\pi})^2$ and $c=\pi^{-1} (1-\pi)^{1/6} (1+\sqrt{1-\pi})^{4/3}$. We can now derive (\ref{tN2.5}) formally. From (\ref{teq:R.9}), Lemma \ref{tlem:A.1}, and (\ref{teq:A.20}) we learn that we should have
\begin{eqnarray}\label{tN2.10}
t-k = bk + ck^{1/3}(-s)(1 + o(1))\; ,
\end{eqnarray}
for $t \to \infty$ and with $k = Jt + Vst^{1/3}$. A straight forward calculation shows that
this can only be achieved if
\[
(b+1)J = 1 \quad \mbox{and} \quad V(b+1) = c J^{1/3}
\]
leading to the formulae for $J$ and $V$ as presented in Theorem \ref{tsatz:R.1}. Moreover, it is apparent that the remainder $o(1)$ in (\ref{tN2.10}) is of order ${\cal O}\left(t^{-2/3}\right)$.
Clearly, the argument just made does not fully prove Theorem \ref{tsatz:R.1} since additional estimates are needed to deduce the convergence of Fredholm determinants from the convergence of the kernels. For details see \cite[Sect. 3]{Jo1}.

Finally we observe that the linear growth of the mean and the $t^{1/3}$-scaling of the fluctuations of the flux $F_0(t)$ follow directly from condition (\ref{tN2.10}), i.e. from (\ref{teq:R.9}), Lemma \ref{tlem:A.1}, and (\ref{teq:A.20}) without any reference to KPZ theory.

\subsection{Remarks on the universal behavior of orthogonal polynomials}
\label{tsec:A.3}

We have discussed two examples for fluctuations that can be
described by the Tracy-Widom distribution: The largest eigenvalue of
GUE and the particle flux at the origin of dTASEP with step initial
conditions. Both systems can be analyzed by the method of orthogonal
polynomials. The fact that the limiting distribution for the
fluctuations agree in both cases is then a consequence of the fact
that the corresponding sets of orthogonal polynomials, Hermite and
Meixner, have the same asymptotic behavior in a vicinity of their
respective largest zeros after appropriate rescaling. This is no
coincidence. During the past ten years many detailed results on
various types of orthogonal polynomials have become available that
show universal behavior of orthogonal polynomials of large degree on
a local scale. In this section we give a rough description of this
universal behavior and outline a few approaches how such results can
be proved.

Let us assume that the support of the measure of orthogonality $\alpha$ is contained in $\mathbb R$. Then the normalized orthogonal polynomial $q_n$ of degree $n$ with positive leading coefficient, that is defined through
\[
\int\limits_{\mathbb R} q_n(x)q_m(x)\, d\alpha (x) =\delta_{n,m}\; ,
\]
has $n$ simple real roots which we denote by $x_i^{(n)}$. For many measures of orthogonality $\alpha$, and in particular in the case of the so called varying weights which we do not discuss here any further, there exists some natural scaling $x \to \hat{x}$ such that the
counting measures $\frac{1}{n}\sum^n_{i=1} \delta_{\hat{x}_i^{(n)}}$ associated with the rescaled zeros $\hat{x}_i^{(n)}$
converge for $n \to \infty$ to some measure $\mu$ of total mass 1.
The support $S$ of $\mu$ is compact and is always contained in the support of $\alpha$. For example, in the case of Hermite polynomials $d\alpha  = e^{-x^2}\, dx$, the scaling is given by $\hat{x} = x/n^{1/2}$ and the limiting measure of zeros is given by $d\mu = \pi^{-1} \sqrt{2-x^2} \, {\bf 1}_{S}(x) \, dx$ with $S=[-\sqrt{2}, \sqrt{2}]$.

One should note that the scaling $x \to \hat{x}$ as well as the measure of zeros $\mu$ do depend on the measure of orthogonality $\alpha$. In order to explain what is universal about orthogonal polynomials we restrict ourselves to the common case that $S$, the support of $\mu$, consists of a single interval or a finite union of disjoint intervals. It is convenient to describe the behavior of $q_n$ by considering the corresponding functions $\varphi_n$ that are orthonormal with respect to Lebesgue measure. For example, in the case $d\alpha = w(x)dx$ we have $\varphi_n = q_n \sqrt w$. In the situation described above the large $n$ behavior of $\varphi_n(x)$ is generically the following:

\noindent
{\em For $\hat{x}$ outside $S$}: $\varphi_n(x)$ decays at an exponential rate to zero as $n\to\infty$.

\noindent
{\em For $\hat{x}$ in the interior of $S$}: $\varphi_n(x)$ is oscillating rapidly (in $\hat{x}$) and can be described to leading order by the cosinus function with slowly varying amplitude and frequency, which only depend on $\mu$.

\noindent
{\em For $\hat{x}$ close to a boundary point $b$ of $S$}: To leading order $\varphi_n(x)$ can be expressed in terms of special functions. One distinguishes between soft edges ($b$ lies in the interior of the support of $\alpha$) and hard edges ($b$ is also a boundary point of the support of $\alpha$).
In the first case the density of $\mu$ usually vanishes like a square root at $b$ and then the leading order of $\varphi_n(x)$ is described by the Airy function. This is in particular the case for Hermite and Meixner polynomials.
In the case of a hard edge the situation is a bit more complicated. In many cases Bessel functions can be used for asymptotic formulas for $\varphi_n(x)$ (see e.g. \cite{TKuijVanl03, TKMVV04}).

We conclude this section by mentioning a few methods how such asymptotics for $\varphi_n$ can be proved. We will in particular remark on the appearance of the Airy function.

\noindent
{\em I. Differential equations of second order}

We again discuss Hermite polynomials $d\alpha = e^{-x^2}dx$ as a typical example. The corresponding functions
$\varphi_n$ satisfy the second order differential equations
\[
\varphi''_n(x)+(2n+1-x^2)\varphi_n(x)=0\; .
\]
WKB analysis of these differential equations shows that the
oscillatory region $|x| < \sqrt{2n+1}$ is connected with the
exponentially decaying region $|x| > \sqrt{2n+1}$ by Airy functions.
This approach can be applied to a number of classical orthogonal
polynomials that are known to solve linear differential equations of
second order with nice coefficients.

\noindent
{\em II. Representation by contour integrals}

Such representations are known for a number of classical orthogonal polynomials (e.g. for Meixner polynomials, see e.g. \cite[Sect. 5]{Jo1})
and can be analyzed using the method of steepest descent. The appearance of the Airy function can be seen from its integral representation (\ref{tAi.5}) which generically provides a normal form at critical points of higher degeneracy.

\noindent
{\em III. Riemann-Hilbert problems}

The characterization of orthogonal polynomials as unique solutions of certain matrix Riemann-Hilbert problems (see \cite{TDeift99} and references therein)
works in principle for all types of weights and opens in particular the way to analyze non-classical orthogonal polynomials. Here the limiting measure of the zeros $\mu$, which can also be defined as the unique minimizer of a variational problem, yields the key to the asymptotic analysis. In the neighborhoods of boundary points of the support of $\mu$ at which the density
vanishes like a square root -- this is the generic case for a soft edge --
Airy functions arise naturally. This method for the asymptotic analysis of orthogonal polynomials was first
carried out in \cite{TDKMVZ}.
The method works best in the class of analytic weights, but progress has recently been made for weights
that have only a finite number of derivatives \cite{TMcLaughlinMiller08}. Orthogonal polynomials with respect to discrete measures
have been analyzed by Riemann-Hilbert techniques in \cite{TBKMM}.

\noindent
{\em IV. Reproducing kernels}

In recent years universality results for orthogonal polynomials, in particular results on the reproducing kernel (cf. Lemma \ref{tlem:A.1}), have been substantially generalized (see \cite{TLubinsky09}
and references therein). A very nice view on universality has been introduced in \cite{TLubinsky08} where classical results on reproducing kernels for entire functions of exponential type are being used (see \cite{TLevinLubinsky09} for the Airy kernel).

\section{KPZ-universality revisited}

\setcounter{footnote}{0}

\label{Universality}

In view of the universality conjecture formulated in
Sect.~\ref{KPZ_conjecture}, one expects that the results
derived in the preceding sections for a very special case --
the dTASEP with step initial conditions -- should carry over,
in a quantitative sense, to a much broader class of models.
The first explicit demonstration  of this idea
was presented by Pr\"ahofer and Spohn in a series of papers
\cite{Prahofer00b,Prahofer00a,Prahofer01}, where an
alternative and independent route linking Ulam's problem to growth
models was established\footnote{Another link between the two classes
  of problems was found by Majumdar and Nechaev \cite{Majumdar2004,Majumdar2007}. }.

The starting point is the one-dimensional polynuclear growth model (PNG),
an interacting particle system on the real line, in which particles
(antiparticles) move deterministically at unit speed to the right (left),
annihilate upon colliding, and are created in pairs according to a
two-dimensional Poisson process in space and time \cite{Krug89}.
Via the random set of particle creation events the model can be mapped
onto the problem of the longest increasing subsequence of a random
permutation, which in turn provides a link to the
Tracy-Widom distribution \cite{Baik99}.
For the case of a droplet growing from a seed (case III of
Sect.~\ref{KPZ_conjecture}), Pr\"ahofer and Spohn show that the resulting
fluctuation  distribution is
identical (under the rescaling prescribed by KPZ theory) to that obtained
by Johansson for the dTASEP.

\begin{figure}
\begin{center}
% \resizebox{16cm}{!}
\includegraphics[width=0.4\textwidth,angle=-90]{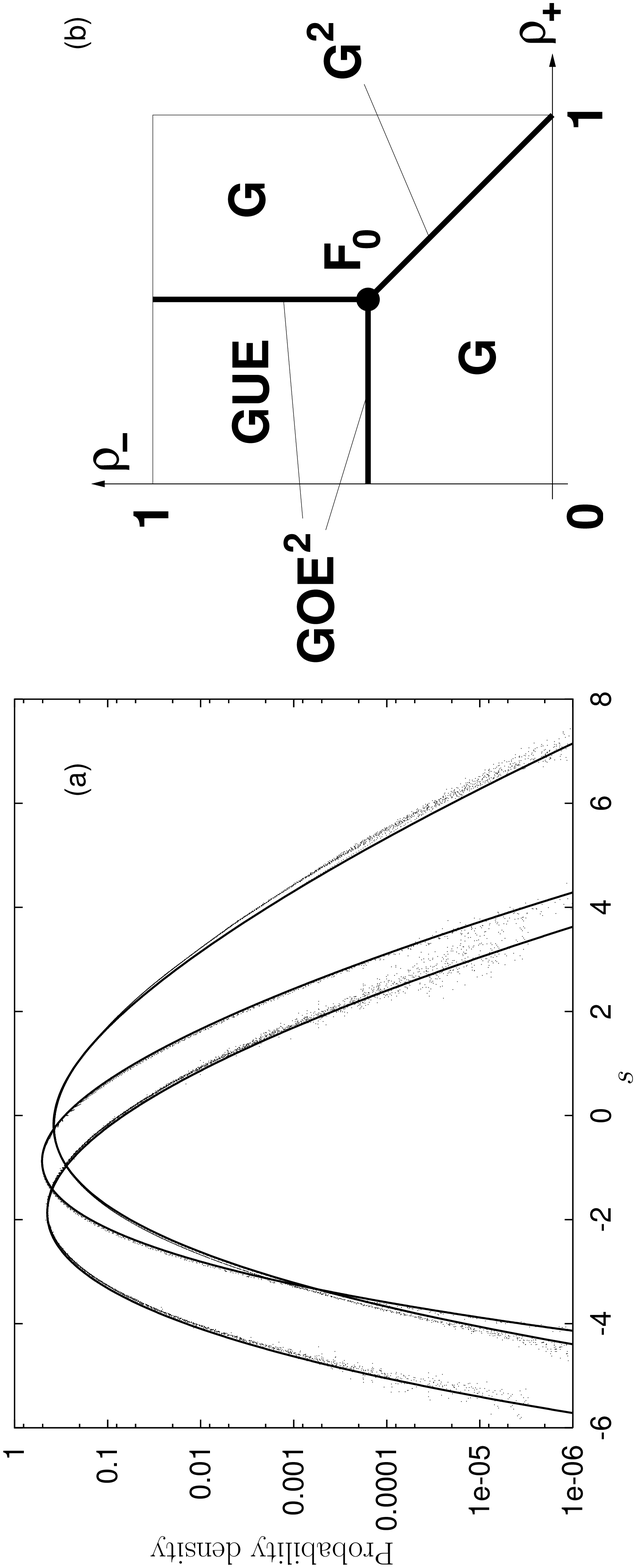}
\end{center}
\caption{(a) The densities of the three universal distribution functions
$TW_2$, $TW_1$ and $F_0$ (from left to right).
Discrete points show simulation results
for the PNG model. Reprinted with permission from
\cite{Prahofer00a}. Copyright 2000 by the American Physical Society. (b) Phase
diagram for the distribution of current fluctuations in the TASEP with
Bernoulli step initial conditions. Here G denotes the Gaussian distribution and
G$^2$ (GOE$^2$) is the distribution of the maximum of two
independent Gaussian ($TW_1$) random variables. 
Initial particle densities are $\rho_L = \rho_-$ to the left and
$\rho_R = \rho_+$ 
to the right of the origin.
Reprinted from \cite{Prahofer01}
with kind permission of Springer Science and Business Media.}
\label{3functions}
\end{figure}

Moreover, by imposing suitable boundary conditions \cite{Baik00} and
symmetry relations \cite{TBaRa01} on the set of Poisson points,
the cases of flat and rough initial conditions (case I and II of
Sect.~\ref{KPZ_conjecture}) can be handled as well \cite{Prahofer00a}.
For the flat initial condition (case I) the fluctuations are governed
by the GOE distribution $TW_1$, while for the rough
initial condition (case II) a new distribution $F_0$ emerges which so
far does not have an interpretation in terms of random matrix theory
\cite{Baik00}. The three distributions are depicted in
Fig.~\ref{3functions}(a).

Since the fundamental works of Johansson, Pr\"ahofer and Spohn the field
has developed rapidly, and it is impossible to do
justice to the new results in the framework of these
lecture notes. In the following subsections we therefore restrict ourselves
to briefly outlining the most important directions of research,
providing the interested reader with a few key references along which
recent advances can be traced.

\subsection{The Pr\"ahofer-Spohn conjecture and the ASEP}
\label{T8.1}

Based on the universality hypothesis, Pr\"ahofer and Spohn
translated the results obtained for the PNG model into a conjecture
for the fluctuations of the particle current through the origin for
the TASEP with a general step initial condition (\ref{step}), where particles are placed to the left (right)
of the origin according to a Bernoulli measure with density $\rho_L$ ($\rho_R$) 
\cite{Prahofer01}. The fluctuation phase diagram in the plane of
the boundary densities is shown
in Fig.~\ref{3functions}(b). The overall features of
the diagram can be understood from hydrodynamics. First, the
Johansson result obtained at $\rho_L = 1$, $\rho_R = 0$ is seen to
extend throughout the region $\rho_L > 1/2 > \rho_R$. As explained
in Sect.~\ref{KPZ_conjecture}, this reflects the fact that the
density profile near the origin is independent of the boundary
densities in this case. For $\rho_L < 1/2$  and $\rho_R > 1/2$ the
application of the hydrodynamic formulae (\ref{wave}) [for $\rho_L >
\rho_R$] and (\ref{shockspeed}) [for $\rho_L < \rho_R$] show that
the density at the origin becomes $\rho_L$  and $\rho_R$,
respectively. In these cases the intrinsic current fluctuations are
masked by the initial fluctuations drifting across the origin,
leading to simple Gaussian statistics (regions marked G in the
diagram). The line $\rho_R + \rho_L = 1$, $\rho_L < \rho_R$, is
special, because there the shock speed (\ref{shockspeed}) vanishes
and the density at the origin shifts randomly between $\rho_L$ and
$\rho_R$. As a consequence, the current is distributed as the
maximum of two independent Gaussian random variables (denoted by
G$^2$ in the figure). Similarly, along the lines $\rho_L = 1/2$,
$\rho_R < 1/2$, and $\rho_R = 1/2$, $\rho_L > 1/2$, the distribution
is that of the maximum of two independent variables drawn from
$TW_1$. Finally, at the point $\rho_L = \rho_R = 1/2$ we have case
II behavior governed by the distribution $F_0$.

A proof of the Pr\"ahofer-Spohn conjecture for the TASEP was
recently presented by Ben Arous and Corwin \cite{TBeCo09} (see also
\cite{TNaSa04,TBaBePe05,Ferrari2006} for earlier partial results). Moreover, in a
remarkable series of papers Tracy and Widom have been able to
generalize these results to the (partially) asymmetric exclusion
process \cite{TTracyWidom08a,TTracyWidom08b,TTracyWidom09a,TTracyWidom09b,TTracyWidom09c,TTracyWidom09d,TTracyWidom10}.
The generalization is highly nontrivial, because the ASEP for
general $q$ is not a determinental process \cite{TTracyWidom09c}, and it requires a
novel set of techniques based on the Bethe ansatz \cite{TSchutz97,TRaSchu05,Gollinelli2006}.

\subsection{Spatio-temporal scaling}
\label{T8.2}

We have seen in Sect.~\ref{KPZ_conjecture} that the essence of the
KPZ conjecture is the universality of height fluctuations when
viewed on the appropriate scales defined by the height rescaling
(\ref{heightscaled}) and the correlation length
(\ref{correlation_length}). In other words, once the average growth
shape has been subtracted, one expects that the rescaled
fluctuations
$$
\bar{h}_t(y) \equiv (A^2 \vert \lambda \vert t)^{-1/3} h(y (A \lambda^2 t^2)^{1/3} ,t)
$$
converge for $t \to \infty$ to a universal stochastic process
${\cal{A}}(y)$, whose single-point distribution is one of the random
matrix distributions discussed above. The process ${\cal{A}}(y)$
was first explicitly characterized by Pr\"ahofer and Spohn for the
PNG-model in the droplet geometry (case III  of
Sect.~\ref{KPZ_conjecture}), who named it the \textit{Airy process}
\cite{TPrSp02}. Subsequently Sasamoto identified the analogous
process for the case of flat initial conditions (case I)
\cite{TSasamoto05}. In line with the nomenclature used to
designate the corresponding single-point distributions ($TW_2$ for
case III and $TW_1$ for case I), the two processes are now called
Airy$_2$ and Airy$_1$ processes, respectively \cite{Borodin07,TBoFeSa08,TBoFe08,TFerrari08b}.
Whereas the Airy$_2$ process has a natural interpretation in the
random matrix context as the motion of the largest eigenvalue
in GUE matrix diffusion (Dyson's Brownian motion), the
corresponding relation does not hold for the Airy$_1$ process
\cite{TBoFePr08}. The corresponding process for initial
conditions with stationary roughness (case II) was studied in
\cite{Baik2010} (see also \cite{Imamura2004}). 

In recent work Ferrari and collaborators have extended the analysis
to include correlations between height fluctuations at different
times $t$ and $t'>t$. As for the scaling of the height fluctuations
themselves, the characteristics of the hydrodynamic equation play a
special role for the decay of correlations. Whereas the decay along
generic space-time directions is governed by the correlation length
(\ref{correlation_length}), which is of order $t^{2/3}$, along the
characteristics the decorrelation time at time $t$ is set by $t$
itself, which implies a much slower decay
\cite{TFerrari08a,TCoFePe10a,TCoFePe10b}. This is in accordance with
KPZ phenomenology, which predicts that such correlations should
decay as $(t/t')^{\bar{\lambda}}$ with a universal autocorrelation
exponent $\bar{\lambda}$ \cite{Krech1997,Kallabis1999}. In contrast
to the scaling exponents of single-point height fluctuations
introduced in Sect.~\ref{KPZ_conjecture}, the autocorrelation
exponent depends explicitly on the growth geometry: For a flat
initial condition (case I) $\bar{\lambda} = 1$, whereas for a curved
cluster (case III) $\bar{\lambda} = 1/3$ \cite{Singha2005}.

\subsection{KPZ-scaling at large}
\label{T8.3}

The results described so far in this section were based on a small
set of exactly solvable models, the (T)ASEP's and the PNG model. On
the other hand, KPZ universality is expected to hold for a much
broader class of interacting particle systems and growth models
which is limited only by the requirement of local, stochastic
transition rules and a nonlinear dependence of the particle current
(or growth rate) on the particle density (or surface slope) [see
Sect.~\ref{KPZ_conjecture}]. It is therefore gratifying that the
class of models for which KPZ universality has been rigorously
established -- mostly in the sense of finding the exact scaling
exponents governing the order of fluctuations -- has been greatly
expanded in recent years. This has required the development of new
techniques that are purely probabilistic in nature and do not rely
on the specific analytic structure of the exactly solvable models.
The first result of this type was obtained in \cite{TCaGr06} for the
Hammersley process, an exlusion-type process in continuous space
which is closely related to the Ulam problem. Similar methods were
subsequently applied to a variety of interacting particle systems
\cite{TQuVa07,TQuVa08,TBaSe,TBaSe09}, including a class of zero
range processes\footnote{Zero range processes were introduced by
Spitzer \cite{Spitzer1970} and have been extensively studied in the
physics literature, see \cite{Evans2005}.} with general jump rates
that have a non-decreasing, concave dependence on the number of
particles \cite{TBaKoSe}. This constitutes a major step on the way
to proving KPZ universality in the broadest sense.

\subsection{The universality class of the KPZ equation}
\label{T8.4}

Ironically, although tremendous advances in the analysis of different representatives of the KPZ universality class
were achieved over the past decade, the one-dimensional KPZ equation (\ref{KPZ}) itself remained rather poorly understood.
We noted already
that the KPZ equation is mathematically ill-posed because of the highly
singular white noise term, and some regularization is needed to make it amenable to rigorous analysis. This can be
done by spatial discretization or by constructing the equation through a scaling limit from an asymmetric
exclusion process with weak asymmetry \cite{TBeGi97,TSaSp10a}. Both approaches have recently been used to prove the
correct order of fluctuations in the KPZ equation \cite{TSaSp09,TBaQuSe10} as well as refined
universality in the sense of the Tracy-Widom distribution \cite{TSaSp10b,TSaSp10c,Amir2010}. Thus it has finally been
established, as it were, that the KPZ equation belongs to its own universality class.

In another line of work an
independent, non-rigorous approach to establishing Tracy-Widom
universality has been developed which is based on applying the replica
method to the path weight $Z(x,t)$ in the Hopf-Cole
transformed equation (\ref{Z(x,t)}). In this approach one computes
moments $\mathbb{E}(Z^n)$ with respect to the stochastic force
$\zeta(x,t)$, which results in a
problem of $n$ bosonic, quantum-mechanical particles
interacting through an attractive $\delta$-function potential
\cite{Kardar1987b,HalpinHealy1995}. While the ground-state energy and
wave function of this quantum system 
been known for a long time, the recent works
\cite{Calabrese2010,Dotsenko2010}
have succeeded in summing
over the full spectrum of excited states of the many-body Hamiltonian.

\begin{figure}
\begin{center}
\includegraphics[width=0.9\textwidth]{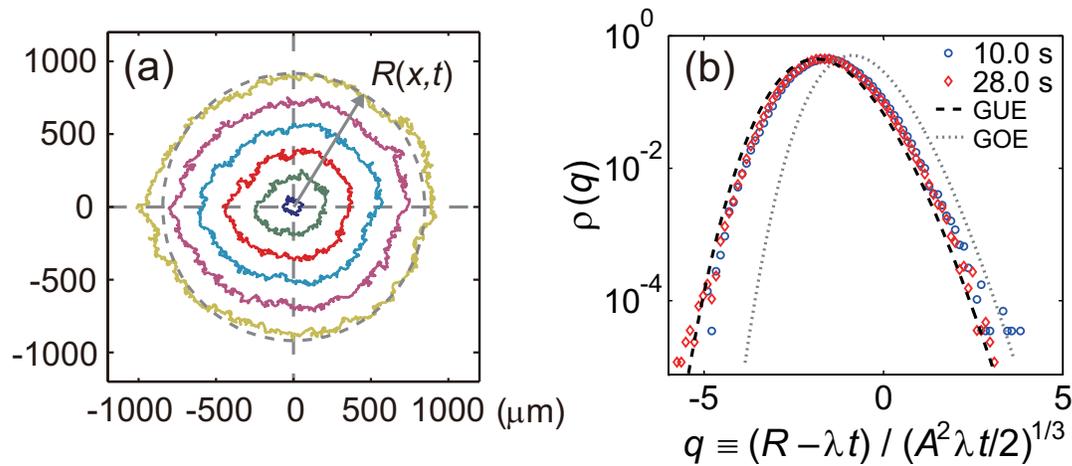}
\end{center}
\caption{(Color online) Experimental demonstration of KPZ universality in a thin film
  of turbulent liquid crystal. (a) Outlines of a growing turbulent
droplet at time intervals of 5 s. (b) Measured distribution of
shape fluctuations in comparison to the GUE prediction. Reprinted with permission from
\cite{Takeuchi2010}. Copyright 2010 by the American Physical Society.}
\label{Kazumasa}
\end{figure}

\subsection{Experiments}
\label{T8.5}

Despite the wide applicability of the KPZ theory to a large class of stochastic growth
models, experimental signatures of KPZ scaling in the real physical
world have proven to be surprisingly elusive. Early efforts focused on the investigation of
growth-induced (two-dimensional) surface roughness of crystals and thin solid films
\cite{Krim1995}. However, detailed consideration of the
physical processes governing such growth experiments has revealed that
they typically operate in regimes where KPZ asymptotics is practically
out of reach \cite{Krug1997,Michely2004}.
To date, the most thoroughly
studied experimental system that displays KPZ scaling is the slow, flameless
combustion of paper. By imaging and analyzing the one-dimensional smoldering front,
the $t^{1/3}$-scaling of fluctuations \cite{Maunuksela1997} as well as
the non-Gaussian shape of the corresponding probability distribution \cite{Miettinen2005}
were demonstrated.

Very recently, a new experimental system involving different phases
of driven, turbulent liquid crystal films became available, in which
the refined universality predictions of the theory can be tested
with unprecedented precision \cite{Takeuchi2010}. In Fig.~\ref{Kazumasa} we show a
series of snapshots of the growing turbulent droplet along with the
experimentally determined distribution of shape fluctuations. Under
the rescaling prescribed by KPZ theory, the predicted universal GUE
distribution is seen to emerge from the data without any adjustable
parameters.

\section{Integrability and Universality}

The central model discussed in our paper is the totally asymmetric
simple exclusion process (TASEP). This model has two features that
motivate our choice. Firstly, it belongs to the class of stochastic
interacting particle systems that are useful in the study of
transport phenomena in nonequilibrium systems. We are particularly
interested in the description of fluctuations around the mean
behavior which is governed by a deterministic evolution equation. In
general it is difficult to obtain such detailed information, mainly
because the interactions between the particles destroy stochastic
independence and the classical central limit theorem cannot be
applied. However, and this is the second feature, TASEP is an
exactly solvable model and the fluctuations can be analyzed by a
series of beautiful and non-obvious observations. We have seen by
explicit calculation that the fluctuations of the flux are described
by a formula that is structurally the same as the formula for the
fluctuations of the largest eigenvalue of matrices from the Gaussian
Unitary Ensemble (GUE). This provides one explanation for the much
celebrated link to Random Matrix Theory. There are also links
to the theory of integrable systems (see e.g. \cite{TDeift99, TMoerbeke02, TMoerbeke08}) that we have not explained at all and that we
mention here only in order to give an indication why TASEP is sometimes called an integrable model, even though there is no differential equation to integrate in sight.

A common feature of integrable systems is that their delicate
mathematical structure immediately breaks down if the model is
changed ever so slightly. One may think that this limits the
interest in the corresponding results. However, the recurring
experience with integrable systems has been, that even though the
method of proof is not applicable if the model is changed, the results may
persist. In our context a nice example of this principle of
universality is the recent work of Tracy and Widom where it is shown
by very different methods that a number of results for TASEP also
hold for ASEP. Universality results are also available for large
classes of matrix ensembles (see \cite{TDeiftGioev09, TErdos10} and references therein).
Despite these results, the question of universality is a subtle one
as can be seen for example from the
correlations of the fluctuations (cf. \cite{TBoFePr08} and Sect.~\ref{T8.2}). It remains
a great challenge to understand the realm of validity of the various
laws that have been established for the specific integrable models.

\section*{Acknowledgements}
This work was supported and inspired by SFB/TR 12 \textit{Symmetries
and universality in mesoscopic systems}. We are grateful to H. Spohn,
P.L. Ferrari, and to an unknown referee for valuable remarks, and to
M. Pr\"ahofer and K.A. Takeuchi for providing figures.


\section*{References}

\begin{thebibliography}{100}

\bibitem{Aldous1999}
D.~Aldous and P.~Diaconis.
\newblock Longest increasing subsequences: from patience
sorting to the {B}aik-{D}eift-{J}ohansson theorem.
\newblock {\em Bull. Amer. Math. Soc.}, 36:413--432, 1999.

\bibitem{Amar1992}
J.~G. Amar and F.~Family.
\newblock Universality in surface growth - scaling functions and amplitude
  ratios.
\newblock {\em Phys. Rev. A}, 45:5378--5393, 1992.

\bibitem{Amir2010}
G.~Amir, I.~Corwin, and J.~Quastel.
\newblock Probability distribution of the free energy of the continuum directed
  random polymer in 1+1 dimensions.
\newblock {\em arXiv:math.PR}, 1003.0443v2, 2010 (to appear in
Comm. Pure Appl. Math.).

\bibitem{TBeCo09}
G.~Ben Arous and I.~Corwin.
\newblock Current fluctuations for {TASEP}: A proof of the {P}r\"ahofer-{S}pohn
  conjecture.
\newblock {\em arXiv:math.PR}, 0905.2993, 2009 (to appear in Ann. Prob.).

\bibitem{TBaBePe05}
J.~Baik, G.~Ben~Arous, and S.~P{\'e}ch{\'e}.
\newblock Phase transition of the largest eigenvalue for nonnull complex sample
  covariance matrices.
\newblock {\em Ann. Probab.}, 33(5):1643--1697, 2005.

\bibitem{Baik99}
J.~Baik, P.~Deift, and K.~Johansson.
\newblock On the distribution of the length of the longest increasing
  subsequence of random permutations.
\newblock {\em J. Am. Math. Soc.}, 12:1119--1178, 1999.

\bibitem{Baik2010}
J.~Baik, P.~L. Ferrari, and S.~P\'ech\'e.
\newblock Limit process of stationary TASEP near the characteristic line.
\newblock {\em Comm. Pure Appl. Math.}, 63:1017--1070, 2010.

\bibitem{TBKMM}
J.~Baik, T.~Kriecherbauer, K.~T.-R. McLaughlin, and P.~D. Miller.
\newblock {\em Discrete orthogonal polynomials}, volume 164 of {\em Annals of
  Mathematics Studies}.
\newblock Princeton University Press, Princeton, NJ, 2007.
\newblock Asymptotics and applications.

\bibitem{Baik00}
J.~Baik and E.~M. Rains.
\newblock Limiting distributions for a polynuclear growth model with external
  sources.
\newblock {\em J. Stat. Phys.}, 100:523--541, 2000.

\bibitem{TBaRa01}
J.~Baik and E.~M. Rains.
\newblock Symmetrized random permutations.
\newblock In {\em Random matrix models and their applications}, volume~40 of
  {\em Math. Sci. Res. Inst. Publ.}, pages 1--19. Cambridge Univ. Press,
  Cambridge, 2001.

\bibitem{TBDS}
J.~Baik, P.~Deift, and T.~Suidan.
\newblock {\em Some combinatorial problems and random matrix theory}.
\newblock (monograph in preparation).

\bibitem{TBaKoSe}
M.~Bal{\'a}zs, J.~Komj{\'a}thy, and T.~Sepp{\"a}l{\"a}inen.
\newblock Microscopic concavity and fluctuation bounds in a class of deposition
  processes.
\newblock {\em arXiv:math.PR}, 0808.1177, 2008.

\bibitem{TBaQuSe10}
M.~Bal{\'a}zs, J.~Quastel, and T.~Sepp{\"a}l{\"a}inen.
\newblock Scaling exponent for the {H}opf-{C}ole solution of {KPZ}/stochastic
  {B}urgers.
\newblock {\em arXiv:math.PR}, 0909.4816, 2010.

\bibitem{Balazs2006}
M.~Bal\'azs, F.~Rassoul-Agha, and T.~Sepp\"al\"ainen.
\newblock The random average process and random walk in a space-time random
  environment in one dimension.
\newblock {\em Commun. Math. Phys.}, 266:499--545, 2006.

\bibitem{TBaSe}
M.~Bal{\'a}zs and T.~Sepp{\"a}l{\"a}inen.
\newblock Order of current variance and diffusivity in the asymmetric simple
  exclusion process.
\newblock {\em Ann. Math.}, 171:1237--1265, 2010.

\bibitem{TBaSe09}
M.~Bal{\'a}zs and T.~Sepp{\"a}l{\"a}inen.
\newblock Fluctuation bounds for the asymmetric simple exclusion process.
\newblock {\em ALEA Lat. Am. J. Probab. Math. Stat.}, 6:1--24, 2009.

\bibitem{TBeGi97}
L.~Bertini and G.~Giacomin.
\newblock Stochastic {B}urgers and {KPZ} equations from particle systems.
\newblock {\em Comm. Math. Phys.}, 183(3):571--607, 1997.

\bibitem{Bertini2007}
L.~Bertini, A.~De Sole, D.~Gabrielli, G.~Jona-Lasinio, and C.~Landim.
\newblock Stochastic interacting particle systems out of equilibrium.
\newblock {\em J. Stat. Mech.: Theory Exp.}, P07014, 2007.

\bibitem{Binder1994}
P.~M. Binder, M.~Paczuski, and M.~Barma.
\newblock Scaling of fluctuations in one-dimensional interface and hopping
  models.
\newblock {\em Phys. Rev. E}, 49(2):1174--1181, 1994.

\bibitem{Blythe2007}
R.~A. Blythe and M.~R. Evans.
\newblock Nonequilibrium steady states of matrix-product form: a solver's
  guide.
\newblock {\em Journal of Physics A}, 40:R333--R441, 2007.

\bibitem{TBornemann09}
F.~Bornemann.
\newblock On the numerical evaluation of distributions in random
matrix theory: a review.
\newblock {\em arXiv:math.PR}, 0904.1581, 2009 (to appear in Markov
Proc. Rel. Fields).

\bibitem{TBoFePr08}
F.~Bornemann, P.~L. Ferrari, and M.~Pr{\"a}hofer.
\newblock The {${\rm Airy}_1$} process is not the limit of the largest
  eigenvalue in {GOE} matrix diffusion.
\newblock {\em J. Stat. Phys.}, 133(3):405--415, 2008.

\bibitem{TBoFe08}
A.~Borodin and P.~L. Ferrari.
\newblock Large time asymptotics of growth models on space-like paths. {I}.
  {P}ush{ASEP}.
\newblock {\em Electron. J. Probab.}, 13:no. 50, 1380--1418, 2008.

\bibitem{Borodin07}
A.~Borodin, P.~L. Ferrari, M.~Pr\"ahofer, and T.~Sasamoto.
\newblock Fluctuation properties of the {TASEP} with periodic initial
  configuration.
\newblock {\em J. Stat. Phys.}, 129(5/6):1055--1080, 2007.

\bibitem{TBoFeSa08}
A.~Borodin, P.~L. Ferrari, and T.~Sasamoto.
\newblock Transition between {${\rm Airy}_1$} and {${\rm Airy}_2$} processes
  and {TASEP} fluctuations.
\newblock {\em Comm. Pure Appl. Math.}, 61(11):1603--1629, 2008.

\bibitem{Bray1994}
A.~J. Bray.
\newblock Theory of phase-ordering kinetics.
\newblock {\em Advances in Physics}, 43:357--459, 1994.

\bibitem{Calabrese2010}
P.~Calabrese, P.~Le~Doussal, and A.~Rosso.
\newblock Free energy distribution of the directed polymer at high temperature.
\newblock {\em EPL}, 90:20002, 2010.

\bibitem{TCaGr06}
E.~Cator and P.~Groeneboom.
\newblock Second class particles and cube root asymptotics for {H}ammersley's
  process.
\newblock {\em Ann. Probab.}, 34(4):1273--1295, 2006.

\bibitem{Chowdhury2000}
D.~Chowdhury, L.~Santen, and A.~Schadschneider.
\newblock Statistical physics of vehicular traffic and some related systems.
\newblock {\em Physics Reports}, 329:199--329, 2000.

\bibitem{Cole1951}
J.D. Cole.
\newblock On a quasi-linear parabolic equation occurring in aerodynamics.
\newblock {\em Quart. Appl. Math.}, 9:225--236, 1951.

\bibitem{TCoFePe10b}
I.~Corwin, P.~L. Ferrari, and S.~P\'ech\'e.
\newblock Limit processes for {TASEP} with shocks and rarefaction fans.
\newblock {\em J. Stat. Phys.}, 140:232--267, 2010.

\bibitem{TCoFePe10a}
I.~Corwin, P.~L. Ferrari, and S.~P\'ech\'e.
\newblock Universality of slow decorrelation in {KPZ} growth.
\newblock {\em arXiv:math.PR}, 1001.5345, 2010.

\bibitem{TDeift99}
P.~Deift.
\newblock {\em Orthogonal polynomials and random matrices: a
  {R}iemann-{H}ilbert approach}, volume~3 of {\em Courant Lecture Notes in
  Mathematics}.
\newblock New York University Courant Institute of Mathematical Sciences, New
  York, 1999.

\bibitem{TDeift07}
P.~Deift.
\newblock Universality for mathematical and physical systems.
\newblock In {\em International {C}ongress of {M}athematicians. {V}ol. {I}},
  pages 125--152. Eur. Math. Soc., Z\"urich, 2007.

\bibitem{TDeiftGioev09}
P.~Deift and D.~Gioev.
\newblock {\em Random matrix theory: invariant ensembles and universality},
  volume~18 of {\em Courant Lecture Notes in Mathematics}.
\newblock Courant Institute of Mathematical Sciences, New York, 2009.

\bibitem{TDKMVZ}
P.~Deift, T.~Kriecherbauer, K.~T.-R. McLaughlin, S.~Venakides, and X.~Zhou.
\newblock Strong asymptotics of orthogonal polynomials with respect to
  exponential weights.
\newblock {\em Comm. Pure Appl. Math.}, 52(12):1491--1552, 1999.

\bibitem{DeMasi1989}
A.~DeMasi, E.~Presutti, and E.~Scacciatelli.
\newblock The weakly asymmetric simple exclusion process.
\newblock {\em Annales de l'institut Henri Poincar{\'e} (B) Probabilit{\'e}s et
  Statistiques}, 25:1--38, 1989.

\bibitem{Derrida2007}
B.~Derrida.
\newblock Non-equilibrium steady states: fluctuations and large deviations of
  the density and of the current.
\newblock {\em J. Stat. Mech.: Theory Exp.}, P07023, 2007.

\bibitem{Derrida1993}
B.~Derrida, S.~A. Janowsky, J.~L. Lebowitz, and E.~R. Speer.
\newblock Exact solution of the totally asymmetric exclusion process: Shock
  profiles.
\newblock {\em J. Stat. Phys.}, 73:813--842, 1993.

\bibitem{Derrida1991}
B.~Derrida, J.~L. Lebowitz, E.~R. Speer, and H.~Spohn.
\newblock Dynamics of an anchored {T}oom interface.
\newblock {\em J. Phys. A}, 24(20):4805--4834, 1991.

\bibitem{Devillard1992}
P.~Devillard and H.~Spohn.
\newblock Universality class of interface growth with reflection symmetry.
\newblock {\em J. Stat. Phys.}, 66(3-4):1089--1099, 1992.

\bibitem{Dotsenko2010}
V.~Dotsenko.
\newblock Bethe ansatz derivation of the {T}racy-{W}idom distribution for
  one-dimensional directed polymers.
\newblock {\em EPL}, 90:20003, 2010.

\bibitem{TErdos10}
L.~Erd{\H{o}}s.
\newblock Universality of {W}igner random matrices: a survey of recent results.
\newblock {\em arXiv:math-ph}, 1004.0861, 2010.

\bibitem{Evans2005}
M.~R. Evans and T.~Hanney.
\newblock Nonequilibrium statistical mechanics of the zero-range process and
  related models.
\newblock {\em Journal of Physics A}, 38:R195--R239, 2005.

\bibitem{Ferrari1992}
P.~A. Ferrari.
\newblock Shock fluctuations in asymmetric simple exclusion.
\newblock {\em Prob. Theory Rel. Fields}, 91:81--101, 1992.

\bibitem{Ferrari1998}
P.~A. Ferrari and L.~R.~G. Fontes.
\newblock Fluctuations of a surface submitted to a random average process.
\newblock {\em Electron. J. Probab.}, 3(6):34, 1998.

\bibitem{Ferrari1991}
P.~A. Ferrari, C.~Kipnis, and E.~Saada.
\newblock Microscopic structure of travelling waves in the asymmetric simple
  exclusion process.
\newblock {\em Annals of Probability}, 19:226--244, 1991.

\bibitem{Ferrari2006}
P.~L. Ferrari and H.~Spohn.
\newblock Scaling limit for the space-time covariance of the
stationary totally asymmetric simple exclusion process.
\newblock {\em Commun. Math. Phys.}, 265:1--44, 2006.

\bibitem{TFerrari08a}
P.~L. Ferrari.
\newblock Slow decorrelations in {K}ardar-{P}arisi-{Z}hang growth.
\newblock {\em J. Stat. Mech.: Theory Exp.}, P07022, 2008.

\bibitem{TFerrari08b}
P.~L. Ferrari.
\newblock The universal {${\rm Airy}_1$} and {${\rm Airy}_2$} processes in the
  totally asymmetric simple exclusion process.
\newblock In {\em Integrable systems and random matrices}, volume 458 of {\em
  Contemp. Math.}, pages 321--332. Amer. Math. Soc., Providence, RI, 2008.

\bibitem{Forster1977}
D.~Forster, D.~R. Nelson, and M.~J. Stephen.
\newblock Large-distance and long-time properties of a randomly stirred fluid.
\newblock {\em Phys. Rev. A}, 16:732 -- 749, 1977.

\bibitem{Frey1996}
E.~Frey, U.~C. T\"auber, and T.~Hwa.
\newblock Mode-coupling and renormalization group results for the noisy
  {B}urgers equation.
\newblock {\em Phys. Rev. E}, 53:4424--4438, 1996.

\bibitem{Gollinelli2006}
O.~Golinelli and K.~Mallick.
\newblock The asymmetric simple exclusion process: an integrable model for
  non-equilibrium statistical mechanics.
\newblock {\em J. Phys. A: Math. Gen.}, 39(41):12679--12705, 2006.

\bibitem{Hager2001}
J.~Hager, J.~Krug, V.~Popkov, and G.~M. Sch{\"u}tz.
\newblock Minimal current phase and universal boundary layers in driven
  diffusive system.
\newblock {\em Phys. Rev. E}, 63:056110, 2001.

\bibitem{HalpinHealy1995}
T.~Halpin-Healy and Y.-C. Zhang.
\newblock Kinetic roughening phenomena, stochastic growth, directed polymers
  and all that. {A}spects of multidisciplinary statistical mechanics.
\newblock {\em Physics Reports}, 254:215--414, 1995.

\bibitem{Hopf1950}
E.~Hopf.
\newblock The partial differential equation $u_t + u u_x = \mu u_{xx}$.
\newblock {\em Comm. Pure Appl. Math.}, 3:201--230, 1950.

\bibitem{Huse1985}
D.~A. Huse, C.~L. Henley, and D.~S. Fisher.
\newblock Huse, {H}enley, and {F}isher respond.
\newblock {\em Phys. Rev. Lett.}, 55:2924, 1985.

\bibitem{Imamura2004}
T.~Imamura and T.~Sasamoto.
\newblock Fluctuations of the one-dimensional polynuclear growth model
with external sources.
\newblock {\em Nucl. Phys. B}, 699:503--544, 2004.

\bibitem{Jo1}
K.~Johansson.
\newblock Shape fluctuations and random matrices.
\newblock {\em Comm. Math. Phys.}, 209(2):437--476, 2000.

\bibitem{TJohansson02}
K.~Johansson.
\newblock Non-intersecting paths, random tilings and random matrices.
\newblock {\em Probab. Theory Related Fields}, 123(2):225--280, 2002.

\bibitem{Kallabis1999}
H.~Kallabis and J.~Krug.
\newblock Persistence of {K}ardar-{P}arisi-{Z}hang interfaces.
\newblock {\em Europhys. Lett.}, 45(1):20--25, 1999.

\bibitem{Kardar1987b}
M.~Kardar.
\newblock Replica {B}ethe ansatz studies of two-dimensional interfaces with
  quenched random impurities.
\newblock {\em Nuclear Physics B}, 290:582--602, 1987.

\bibitem{Kardar1986}
M.~Kardar, G.~Parisi, and Y.-C. Zhang.
\newblock Dynamic scaling of growing interfaces.
\newblock {\em Phys. Rev. Lett.}, 56:889 -- 892, 1986.

\bibitem{Kardar1987}
M.~Kardar and Y.-C. Zhang.
\newblock Scaling of directed polymers in random media.
\newblock {\em Phys. Rev. Lett.}, 58:2087--2090, 1987.

\bibitem{Kelly1979}
F.~Kelly.
\newblock {\em Reversibility and stochastic networks}.
\newblock Wiley, New York, 1979.

\bibitem{Kipnis1999}
C.~Kipnis and C.~Landim.
\newblock {\em Scaling limits of interacting particles systems}.
\newblock Springer, Berlin, 1999.

\bibitem{Knu}
D.~E. Knuth.
\newblock Permutations, matrices, and generalized {Y}oung tableaux.
\newblock {\em Pacific J. Math.}, 34:709--727, 1970.

\bibitem{TKonig05}
W.~K{\"o}nig.
\newblock Orthogonal polynomial ensembles in probability theory.
\newblock {\em Probab. Surv.}, 2:385--447 (electronic), 2005.

\bibitem{Krech1997}
M.~Krech.
\newblock Short-time scaling behavior of growing interfaces.
\newblock {\em Phys. Rev. E}, 55:668--679, 1997.

\bibitem{Krim1995}
J.~Krim and G.~Palasantzas.
\newblock Experimental observations of self-affine scaling and kinetic
  roughening at submicron lengthscales.
\newblock {\em Int. J. Mod. Phys. B}, 9(6):599--632, 1995.

\bibitem{Krug1991}
J.~Krug.
\newblock Boundary-induced phase transitions in driven diffusive systems.
\newblock {\em Phys. Rev. Lett.}, 67:1881--1885, 1991.

\bibitem{Krug1997}
J.~Krug.
\newblock Origins of scale invariance in growth processes.
\newblock {\em Advances in Physics}, 46:139--282, 1997.

\bibitem{Krug2000}
J.~Krug.
\newblock Phase separation in disordered exclusion models.
\newblock {\em Brazilian Journal of Physics}, 30:97--104, 2000.

\bibitem{Krug2000a}
J.~Krug and J.~Garc\'{i}a.
\newblock Asymmetric particle systems on $\mathbb{R}$.
\newblock {\em J. Stat. Phys.}, 99(1/2):31--55, 2000.

\bibitem{Krug1998}
J.~Krug and T.~Halpin-Healy.
\newblock Ground-state energy anisotropy for directed polymers in random media.
\newblock {\em Journal of Physics A}, 31:5939--5952, 1998.

\bibitem{Krug1992}
J.~Krug, P.~Meakin, and T.~Halpin-Healy.
\newblock Amplitude universality for driven interfaces and directed polymers in
  random media.
\newblock {\em Phys. Rev. A}, 45:638 -- 653, 1992.

\bibitem{Krug1988}
J.~Krug and H.~Spohn.
\newblock Universality classes for deterministic surface growth.
\newblock {\em Phys. Rev. A}, 38:4271--4283, 1988.

\bibitem{Krug89}
J.~Krug and H.~Spohn.
\newblock Anomalous fluctuations in the driven and damped sine-{G}ordon chain.
\newblock {\em Europhys. Lett.}, 8:219--224, 1989.

\bibitem{Krug1991b}
J.~Krug and H.~Spohn.
\newblock Kinetic roughening of growing surfaces.
\newblock In C.~Godr{\`e}che, editor, {\em Solids far from equilibrium}.
  Cambridge University Press, Cambridge, 1991.

\bibitem{Krug1994}
J.~Krug and L.-H. Tang.
\newblock Disorder-induced unbinding in confined geometries.
\newblock {\em Phys. Rev. E}, 50:104, 1994.

\bibitem{TKMVV04}
A.~B.~J. Kuijlaars, K.~T.-R. McLaughlin, W.~Van~Assche, and M.~Vanlessen.
\newblock The {R}iemann-{H}ilbert approach to strong asymptotics for orthogonal
  polynomials on {$[-1,1]$}.
\newblock {\em Adv. Math.}, 188(2):337--398, 2004.

\bibitem{TKuijVanl03}
A.~B.~J. Kuijlaars and M.~Vanlessen.
\newblock Universality for eigenvalue correlations at the origin of the
  spectrum.
\newblock {\em Comm. Math. Phys.}, 243(1):163--191, 2003.

\bibitem{Lebowitz88}
J.~L. Lebowitz, E.~Presutti, and H.~Spohn.
\newblock Microscopic models of hydrodynamic behavior.
\newblock {\em J. Stat. Phys.}, 51:841--862, 1988.

\bibitem{TLevinLubinsky09}
E.~Levin and D.~S. Lubinsky.
\newblock On the {A}iry reproducing kernel, sampling series, and quadrature
  formula.
\newblock {\em Integral Equations Operator Theory}, 63(3):427--438, 2009.

\bibitem{Liggett1999}
T.~M. Liggett.
\newblock {\em Stochastic interacting systems: contact, voter and exclusion
  processes}.
\newblock Springer, Berlin, 1999.

\bibitem{TLubinsky08}
D.~S. Lubinsky.
\newblock Universality limits in the bulk for arbitrary measures on compact
  sets.
\newblock {\em J. Anal. Math.}, 106:373--394, 2008.

\bibitem{TLubinsky09}
D.~S. Lubinsky.
\newblock Universality limits for random matrices and de {B}ranges spaces of
  entire functions.
\newblock {\em J. Funct. Anal.}, 256(11):3688--3729, 2009.

\bibitem{Majumdar2007}
S.~N. Majumdar.
\newblock Random matrices, the {U}lam problem, directed polymers \& growth
  models, and sequence matching.
\newblock In J.-P. Bouchaud, M.~M\'ezard, and J.~Dalibard, editors, {\em
  Complex systems}. Elsevier, Amsterdam, 2007.

\bibitem{Majumdar2004}
S.N. Majumdar and S.~Nechaev.
\newblock Anisotropic ballistic deposition model with links to the {U}lam
  problem and the {T}racy-{W}idom distribution.
\newblock {\em Phys. Rev. E}, 69:011103, 2004.

\bibitem{Maunuksela1997}
J.~Maunuksela, M.~Myllys, O.-P. K\"ahk\"onen, J.~Timonen, N.~Provatas, M.~J.
  Alava, and T.~Ala-Nissila.
\newblock Kinetic roughening in slow combustion of paper.
\newblock {\em Phys. Rev. Lett.}, 79:1515--1518, 1997.

\bibitem{TMcLaughlinMiller08}
K.~T.-R. McLaughlin and P.~D. Miller.
\newblock The {$\overline{\partial}$} steepest descent method for orthogonal
  polynomials on the real line with varying weights.
\newblock {\em Int. Math. Res. Not. IMRN}, pages Art. ID rnn 075, 66, 2008.

\bibitem{Meakin86}
P.~Meakin, P.~Ramanlal, L.~M. Sander, and R.~C. Ball.
\newblock Ballistic deposition on surfaces.
\newblock {\em Phys. Rev. A}, 34(6):5091--5103, 1986.

\bibitem{TMehta}
M.~L. Mehta.
\newblock {\em Random matrices}, volume 142 of {\em Pure and Applied
  Mathematics (Amsterdam)}.
\newblock Elsevier/Academic Press, Amsterdam, third edition, 2004.

\bibitem{Michely2004}
T.~Michely and J.~Krug.
\newblock {\em Islands, Mounds and Atoms. Patterns and Processes in Crystal
  Growth Far from Equilibrium}, volume~42 of {\em Springer Series in Surface
  Science}.
\newblock Springer, Heidelberg, 2004.

\bibitem{Miettinen2005}
L.~Miettinen, M.~Myllys, J.~Merikoski, and J.~Timonen.
\newblock Experimental determination of {KPZ} height-fluctuation distributions.
\newblock {\em Eur. Phys. J. B}, 46:55--60, 2005.

\bibitem{TNaSa04}
T.~Nagao and T.~Sasamoto.
\newblock Asymmetric simple exclusion process and modified random matrix
  ensembles.
\newblock {\em Nuclear Phys. B}, 699(3):487--502, 2004.

\bibitem{Plischke2006}
M.~Plischke and B.~Bergersen.
\newblock {\em Equilibrium statistical physics}.
\newblock World Scientific, Singapore, 2006.

\bibitem{Plischke87}
M.~Plischke, Z.~R\'acz, and D.~Liu.
\newblock Time-reversal invariance and universality of two-dimensional growth
  models.
\newblock {\em Phys. Rev. B}, 35(7):3485--3495, 1987.

\bibitem{Prahofer00b}
M.~Pr\"ahofer and H.~Spohn.
\newblock Statistical self-similarity of one-dimensional growth processes.
\newblock {\em Physica A}, 279:342--352, 2000.

\bibitem{Prahofer00a}
M.~Pr\"ahofer and H.~Spohn.
\newblock Universal distributions for growth processes in 1+1 dimensions and
  random matrices.
\newblock {\em Phys. Rev. Lett.}, 84:4882--4885, 2000.

\bibitem{Prahofer01}
M.~Pr\"ahofer and H.~Spohn.
\newblock Current fluctuations for the totally asymmetric simple exclusion
  process.
\newblock In Vladas Sidoravicius, editor, {\em In and Out of Equilibrium:
  Probability With a Physics Flavor}. Springer, Berlin, 2002.

\bibitem{TPrSp02}
M.~Pr{\"a}hofer and H.~Spohn.
\newblock Scale invariance of the {PNG} droplet and the {A}iry process.
\newblock {\em J. Statist. Phys.}, 108(5-6):1071--1106, 2002.

\bibitem{TQuVa07}
J.~Quastel and B.~Valk\'o.
\newblock {$t^{1/3}$} {S}uperdiffusivity of finite-range asymmetric exclusion
  processes on {$\mathbb Z$}.
\newblock {\em Comm. Math. Phys.}, 273(2):379--394, 2007.

\bibitem{TQuVa08}
J.~Quastel and B.~Valk{\'o}.
\newblock A note on the diffusivity of finite-range asymmetric exclusion
  processes on {$\mathbb Z$}.
\newblock In {\em In and out of equilibrium. 2}, volume~60 of {\em Progr.
  Probab.}, pages 543--549. Birkh\"auser, Basel, 2008.

\bibitem{TRaSchu05}
A.~R{\'a}kos and G.~M. Sch{\"u}tz.
\newblock Current distribution and random matrix ensembles for an integrable
  asymmetric fragmentation process.
\newblock {\em J. Stat. Phys.}, 118(3-4):511--530, 2005.

\bibitem{Resnick2002}
S.~I. Resnick.
\newblock {\em Adventures in stochastic processes}.
\newblock Birkh{\"a}user, Boston, 2002.

\bibitem{Rost1981}
H.~Rost.
\newblock Non-equilibrium behaviour of a many particle process: Density profile
  and local equilibria.
\newblock {\em Prob. Theory Rel. Fields}, 58:41--53, 1981.

\bibitem{TSagan01}
B.~E. Sagan.
\newblock {\em The symmetric group}, volume 203 of {\em Graduate Texts in
  Mathematics}.
\newblock Springer-Verlag, New York, second edition, 2001.

\bibitem{TSasamoto05}
T.~Sasamoto.
\newblock Spatial correlations of the 1{D} {KPZ} surface on a flat substrate.
\newblock {\em J. Phys. A}, 38(33):L549--L556, 2005.

\bibitem{Sas1}
T.~Sasamoto.
\newblock Fluctuations of the one-dimensional asymmetric exclusion process
  using random matrix techniques.
\newblock {\em J. Stat. Mech. Theory Exp.}, (7):P07007, 31 pp. (electronic),
  2007.

\bibitem{TSaSp09}
T.~Sasamoto and H.~Spohn.
\newblock Superdiffusivity of the {1D} lattice {K}ardar-{P}arisi-{Z}hang
  equation.
\newblock {\em J. Stat. Phys.}, 137:917--935, 2009.

\bibitem{TSaSp10a}
T.~Sasamoto and H.~Spohn.
\newblock The crossover regime for the weakly asymmetric simple exclusion
  process.
\newblock {\em J. Stat. Phys.}, 140:209--231, 2010.

\bibitem{TSaSp10b}
T.~Sasamoto and H.~Spohn.
\newblock Exact height distributions for the {KPZ} equation with narrow wedge
  initial condition.
\newblock {\em Nucl. Phys. B}, 834:523--542, 2010.

\bibitem{TSaSp10c}
T.~Sasamoto and H.~Spohn.
\newblock One-dimensional Kardar-Parisi-Zhang equation: An exact solution and its universality.
\newblock {\em Phys. Rev. Lett.}, 104:230602, 2010.

\bibitem{Schadschneider1993}
A.~Schadschneider and M.~Schreckenberg.
\newblock Cellular-automaton models and traffic flow.
\newblock {\em Journal of Physics A}, 26:L679--L683, 1993.

\bibitem{Schmittmann1995}
B.~Schmittmann and R.~K.~P. Zia.
\newblock {\em Statistical mechnics of driven diffusive systems}.
\newblock Academic, London, 1995.

\bibitem{Schreckenberg1995}
M.~Schreckenberg, A.~Schadschneider, K.~Nagel, and N.~Ito.
\newblock Discrete stochastic models for traffic flow.
\newblock {\em Phys. Rev. E}, 51:2939, 1995.

\bibitem{TSchutz97}
G.~M. Sch{\"u}tz.
\newblock Exact solution of the master equation for the asymmetric exclusion
  process.
\newblock {\em J. Statist. Phys.}, 88(1-2):427--445, 1997.

\bibitem{Schuetz1996}
G.~M. Sch{\"u}tz, R.~Ramaswamy, and M.~Barma.
\newblock Pairwise balance and invariant measures for generalized exclusion
  processes.
\newblock {\em Journal of Physics A}, 29:837--843, 1996.

\bibitem{Seppalainen1999}
T.~Sepp{\"a}l{\"a}inen.
\newblock Existence of hydrodynamics for the totally asymmetric simple
  k-exclusion process.
\newblock {\em Annals of Probability}, 27:361--415, 1999.

\bibitem{Singha2005}
S.~B. Singha.
\newblock Persistence of surface fluctuations in radially growing surfaces.
\newblock {\em J. Stat. Mech.: Theory Exp.}, P08006, 2005.

\bibitem{Spitzer1970}
F.~Spitzer.
\newblock Interaction of {M}arkov processes.
\newblock {\em Adv. Math.}, 5:246--290, 1970.

\bibitem{Spohn1983}
H.~Spohn.
\newblock Long range correlations for stochastic lattice gases in a
  non-equilibrium steady state.
\newblock {\em Journal of Physics A}, 16:4275, 1983.

\bibitem{Spohn1991}
H.~Spohn.
\newblock {\em Large scale dynamics of interacting particles}.
\newblock Springer, Berlin, 1991.

\bibitem{Spohn2006}
H.~Spohn.
\newblock Exact solutions for {KPZ}-type growth processes, random
matrices, and equilibrium shapes of crystals.
\newblock {\em Physica A}, 369:71--99, 2006.

\bibitem{Takeuchi2010}
K.~A. Takeuchi and M.~Sano.
\newblock Universal fluctuations of growing interfaces: Evidence in turbulent
liquid crystals
\newblock {\em Phys. Rev. Lett.}, 104:230601, 2010.

\bibitem{TTracyWidom94}
C.~A. Tracy and H.~Widom.
\newblock Level-spacing distributions and the {A}iry kernel.
\newblock {\em Comm. Math. Phys.}, 159(1):151--174, 1994.

\bibitem{TW2}
C.~A. Tracy and H.~Widom.
\newblock Correlation functions, cluster functions, and spacing distributions
  for random matrices.
\newblock {\em J. Statist. Phys.}, 92(5-6):809--835, 1998.

\bibitem{TTracyWidom02}
C.~A. Tracy and H.~Widom.
\newblock Distribution functions for largest eigenvalues and their
  applications.
\newblock In {\em Proceedings of the {I}nternational {C}ongress of
  {M}athematicians, {V}ol. {I} ({B}eijing, 2002)}, pages 587--596, Beijing,
  2002. Higher Ed. Press.

\bibitem{TTracyWidom08b}
C.~A. Tracy and H.~Widom.
\newblock A {F}redholm determinant representation in {ASEP}.
\newblock {\em J. Stat. Phys.}, 132(2):291--300, 2008.

\bibitem{TTracyWidom08a}
C.~A. Tracy and H.~Widom.
\newblock Integral formulas for the asymmetric simple exclusion process.
\newblock {\em Comm. Math. Phys.}, 279(3):815--844, 2008.

\bibitem{TTracyWidom09a}
C.~A. Tracy and H.~Widom.
\newblock Asymptotics in {ASEP} with step initial condition.
\newblock {\em Comm. Math. Phys.}, 290(1):129--154, 2009.

\bibitem{TTracyWidom09c}
C.~A. Tracy and H.~Widom.
\newblock On {ASEP} with step {B}ernoulli initial condition.
\newblock {\em J. Stat. Phys.}, 137:825--838, 2009.

\bibitem{TTracyWidom09d}
C.~A. Tracy and H.~Widom.
\newblock On the distribution of a second-class particle in the asymmetric
  simple exclusion process.
\newblock {\em J. Phys. A}, 42(42):425002, 6, 2009.

\bibitem{TTracyWidom09b}
C.~A. Tracy and H.~Widom.
\newblock Total current fluctuations in the asymmetric simple exclusion
  process.
\newblock {\em J. Math. Phys.}, 50(9):095204, 4, 2009.

\bibitem{TTracyWidom10}
C.~A. Tracy and H.~Widom.
\newblock Formulas for {ASEP} with two-sided {B}ernoulli initial condition.
\newblock {\em J. Stat. Phys.}, 140:619 -- 634, 2010.

\bibitem{vanBeijeren1985}
H.~van Beijeren, R.~Kutner, and H.~Spohn.
\newblock Excess noise for driven diffusive systems.
\newblock {\em Phys. Rev. Lett.}, 54:2026 -- 2029, 1985.

\bibitem{vanKampen2001}
N.~G. van Kampen.
\newblock {\em Stochastic Processes in Physics and Chemistry}.
\newblock Elsevier, Amsterdam, 2001.

\bibitem{TMoerbeke02}
P.~van Moerbeke.
\newblock Random matrices and permutations, matrix integrals and integrable
  systems.
\newblock {\em Ast\'erisque}, (276):411--433, 2002.
\newblock S{\'e}minaire Bourbaki, Vol. 1999/2000.

\bibitem{TMoerbeke08}
P.~van Moerbeke.
\newblock Nonintersecting {B}rownian motions, integrable systems and orthogonal
  polynomials.
\newblock In {\em Probability, geometry and integrable systems}, volume~55 of
  {\em Math. Sci. Res. Inst. Publ.}, pages 373--396. Cambridge Univ. Press,
  Cambridge, 2008.

\bibitem{Vollmer2002}
J.~Vollmer.
\newblock Chaos, spatial extension, transport, and non-equilibrium
  thermodynamics.
\newblock {\em Physics Reports}, 372:131--267, 2002.

\bibitem{Wolfram1983}
S.~Wolfram.
\newblock Statistical mechanics of cellular automata.
\newblock {\em Reviews of Modern Physics}, 55:601, 1983.

\bibitem{Yaguchi1986}
H.~Yaguchi.
\newblock Stationary measures for an exclusion process on one-dimensional
  lattices with infinitely many hopping sites.
\newblock {\em Hiroshima Mathematical Journal}, 16:449--475, 1986.

\bibitem{Zia2007}
R.~K.~P. Zia and B.~Schmittmann.
\newblock Probability currents as principal characteristics in the statistical
  mechanics of non-equilibrium steady states.
\newblock {\em J. Stat. Mech.: Theory Exp.}, P07012, 2007.

\end{thebibliography}
\end{document}